\documentclass[useAMS,usenatbib]{mn2e}

\usepackage[pdftex]{graphicx}
\usepackage{amssymb,amsmath}
\usepackage{url}
\usepackage{verbatim}

\title[The evolution of star formation in dwarf galaxies]{The slowly evolving role of
environment in a spectroscopic survey of star formation in $M_\ast>5\times 10^8M_\odot$ galaxies since $z\sim 1$ }
\author[C. Greene et al.]
  {Chad R.~Greene$^1$\thanks{Email: cgreene@uwaterloo.ca}, David G.~Gilbank$^{1,2}$\thanks{Email: gilbank@saao.ac.za}, Michael L.~Balogh$^1$, Karl~Glazebrook$^3$, 
  \newauthor 
  Richard G.~Bower$^4$, Ivan K.~Baldry$^5$, George K.T.~Hau$^3$, I.H.~Li$^3$, Pat~McCarthy$^6$
 \\
  $^1$Department of Physics and Astronomy, University of Waterloo, Waterloo, Ontario, Canada N2L 3G1\\
  $^2$ South African Astronomical Observatory, PO Box 9, Observatory, 7935, South Africa\\
  $^3$Centre for Astrophysics and Supercomputing, Swinburne University of Technology, P.O. Box 218, Hawthorn, VIC 3122, Australia\\
  $^4$Institute for Computational Cosmology, Department of Physics, University of Durham, South Road, Durham, DH1 3LE, UK\\
  $^5$Astrophysics Research Institute, Liverpool John Moores University, Twelve Quays House, Egerton Wharf, Birkenhead CH41 1LD, UK\\
  $^6$Carnegie Observatories, 813 Santa Barbara Street, Pasadena, California, 91101 USA}
\date{\today}

\pagerange{\pageref{firstpage}--\pageref{lastpage}} \pubyear{2011}

\def\LaTeX{L\kern-.36em\raise.3ex\hbox{a}\kern-.15em
    T\kern-.1667em\lower.7ex\hbox{E}\kern-.125emX}

\newcommand{\Msun}{{M_{\odot}}}

\def \snframe {\textit{S/N} frame}
\def \snthresh {\textit{S/N} threshold}

\def\msunyr{M$_\odot$ yr$^{-1}$}
\def\oii{[{\sc O\,II}]}
\def\lsim{\mathrel{\hbox{\rlap{\hbox{\lower4pt\hbox{$\sim$}}}\hbox{$<$}}}}
\def\gsim{\mathrel{\hbox{\rlap{\hbox{\lower4pt\hbox{$\sim$}}}\hbox{$>$}}}}

\begin{document}

\label{firstpage}

\maketitle
\begin{abstract}
We present a deep [OII] emission line survey of faint galaxies
($22.5<K_{AB}<24$) in the {\it Chandra} Deep Field South and the
FIRES field.  With these data we measure the star formation rate (SFR) in
galaxies in the stellar mass range
$8.85\lesssim~\log{(M_\ast/\Msun)}~\lesssim9.5$ at $0.62<z<0.885$, to a
limit of SFR$\sim0.1\Msun yr^{-1}$.   The presence of a
massive cluster (MS1054-03) in the FIRES field, and of significant large scale
structure in the CDFS field, allows us to study the environmental
dependence of SFRs amongst this population of low-mass
galaxies.  Comparing
our results with more massive galaxies at this epoch, with our previous survey
(ROLES) at the higher redshift $z\sim 1$, and with SDSS Stripe 82 data, we find no significant evolution of the stellar mass function of
star--forming galaxies between $z=0$ and $z\sim 1$, and no evidence that
its shape depends on environment. 
The correlation between specific star formation
rate (sSFR) and stellar mass at $z\sim0.75$ has a power-law slope of $\beta\sim -0.2$, with evidence for a steeper relation at the lowest masses.  The normalization of this correlation lies as expected between that corresponding to $z\sim1$ and the present
day.  The global SFR density is
consistent with an evolution of the form $(1+z)^2$ over $0<z<1$, with
no evidence for a dependence on stellar mass.
The sSFR of these star--forming galaxies at $z\sim 0.75$ does
not depend upon the density of their local environment.  Considering just high-density environments, the low-mass end of the sSFR-$M_\ast$ relation in our data is steeper than that in Stripe 82 at $z=0$, and shallower than that measured by ROLES at $z=1$.  Evolution of low-mass galaxies in dense environments appears to be more rapid than in the general field.
\end{abstract}
\begin{keywords}
 galaxies: dwarf --- galaxies: evolution --- galaxies: downsizing --- galaxies: environment --- galaxies: general
\end{keywords}
\section{Introduction}
\label{sec:0_introduction}

In recent years, evidence of the bimodal nature of the galaxy
population has been obtained with increasing precision
\citep[e.g.][]{Strateva01_short,Baldry03,Bell+07_short}. Locally, the
galaxy population divides quite cleanly into those which are actively 
star-forming and those in which star-formation has been terminated, or
``quenched". The relative mix of these two populations appears to be
strongly dependent on both environment and stellar mass
\citep[$M_\ast$, e.g.][]{Baldry2006,Peng2010}. 
In particular,
the high mass end of the galaxy stellar mass function 
(GSMF) is dominated by passively evolving galaxies, while the actively 
star--forming population dominates at stellar masses below
$M\sim10^{10}\Msun$\citep[e.g.][]{Pozzetti2009} at
z$\lsim$1.  Observations suggest that star formation is truncated first
in the most massive galaxies
\citep[e.g.][]{Cowie1996,Bundy2006,Pozzetti2009}; however, the stellar
mass function of actively star--forming galaxies itself evolves very
little \citep[e.g.][]{Pozzetti2009,Gilbank2010,Gilbank2010_ROLESII}.

Similarly, in the local universe the relative fraction of star-forming galaxies is
strongly dependent on environment, with  the
densest environments dominated by passive or ``quenched'' galaxies,
and star-forming galaxies preferentially residing in lower density,
``field" environments.  But
the properties of star-forming galaxies themselves have at most a weak dependence on
environment \citep{BB04,STAGES-dust,V+10}.
Recently, several authors have claimed evidence for evolution in this
environment dependence \citep[e.g.][]{Gerke06,zCOSMOS_Cucciati,Bolzonella2010,McGee2011,Patel11,George+11},
with several observations possibly indicating {\it enhanced} star
formation rates (SFRs) in dense regions at 
z$\sim$1, under some circumstances \citep{Elbaz2007,DEEPII_envt_again,I+09,Sobral2011,G+2011}.  All of these effects are relatively subtle, so comparisons
between works are complicated by different definitions of
environment \cite[e.g.][]{Cooper+10,Muldrew+12}, sample selection and choice of estimator (e.g., average
SFR, average specific star formation rate $sSFR=SFR/M_\ast$, SFR of star-forming population), and 
star-formation indicators \citep[e.g.][]{Gilbank2010,Patel11}. Indeed,
previous, apparently contradictory, results may be reconciled when
uniform definitions are adopted \citep{Cooper+10,Sobral2011}. 

\citet{Peng2010} recently presented an illuminating, phenomenological
description encapsulating the environmental and stellar mass dependence
of galaxy activity and suggested that these effects appear to be
entirely separable. 
In their model, the efficiency of environment--driven transformation is
independent of stellar mass and redshift, and the shape of the stellar
mass function (SMF) for star--forming galaxies is universal and
time--independent.  However, their model says nothing about the rate at
which galaxies transform from the star--forming to passive sequence; if
this rate is slow enough, it will be observable as a population of
primarily low--mass galaxies with lower-than-average SFR.   

A direct measurement of this timescale, which would provide important insight into the mechanisms driving this evolution, can be obtained by detecting a population of galaxies currently under the influence of ``environment-quenching''.  The most likely place to find such a signature is amongst low-mass galaxies (for which mass-quenching is ineffective), at redshifts $z>0.5$, when gas fractions and infall rates are high.  
Most spectroscopic surveys at these redshifts are limited to fairly massive galaxies \citep[e.g.][]{Noeske2008,Cooper+10,Bolzonella2010,Patel11,Muzzin2012}.  The Redshift One LDSS-3 Emission Line Survey (hereafter ROLES) was designed to extend this work to lower stellar masses at $z\sim 1$, by searching for emission lines in $K-$selected samples, from fields with very deep imaging \citep{Davies2009}. This was a spectroscopic survey, conducted
using the LDSS-3 instrument on the Magellan (Clay) telescope in Chile.  With a custom made \textit{KG750} filter, 
redshifts and [OII] emission line fluxes were obtained for galaxies at $0.889<z<1.149$ in the
mass range $8.5~<~\log{(M_\ast/M_{\odot})}~<~9.5$. 
\begin{figure*}
  \includegraphics[width=1.0\linewidth]{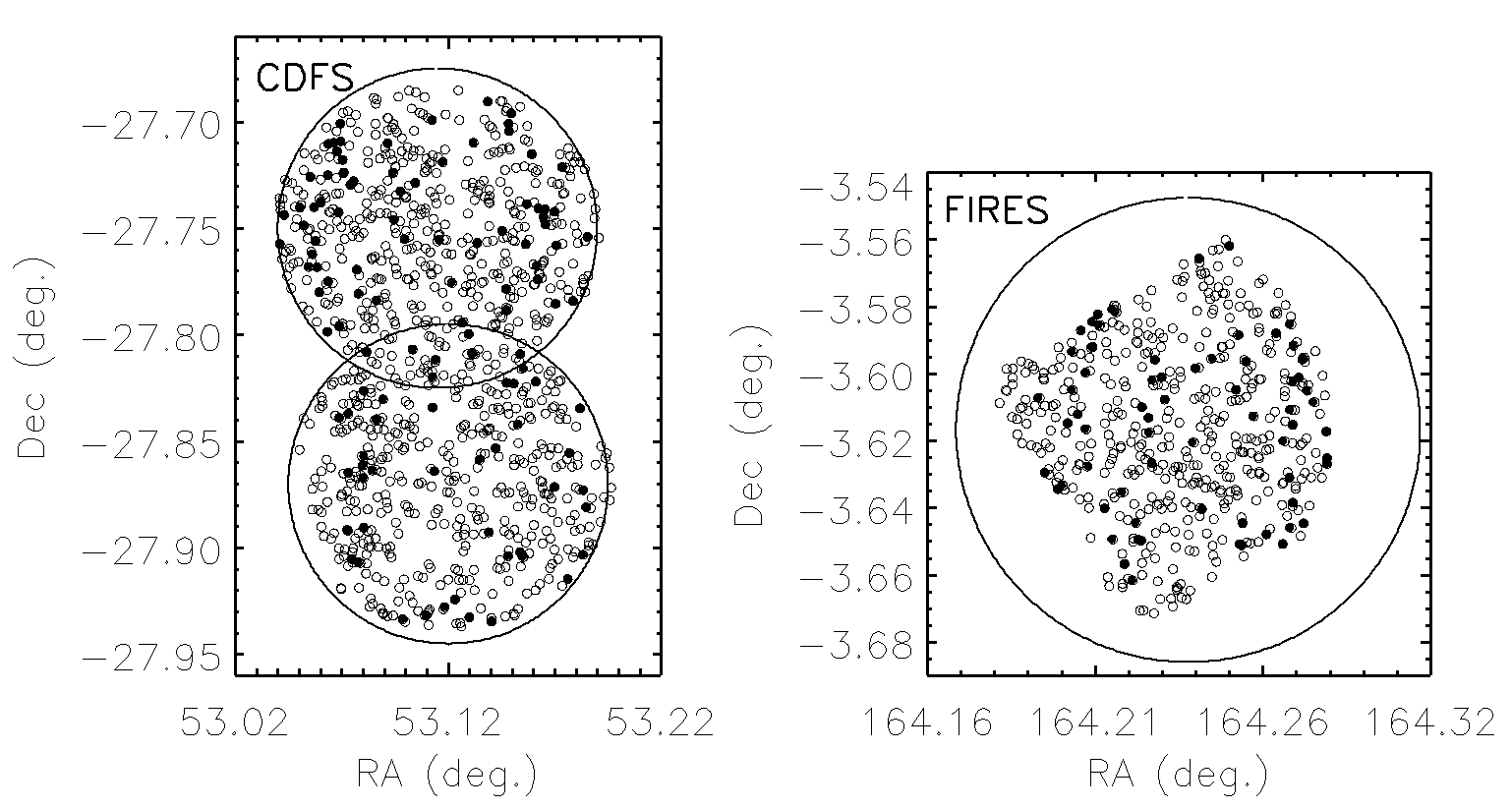}
  \caption{ROLES KG650 pointings in CDFS and FIRES. The ROLES pointings
    in CDFS are centered at $(RA, Dec) = (03^h32^m27.6^s,
    -27^d45^m00^s)$ and $(03^h32^m28.8^s, -27^d52^m12^s)$ while the
    pointing in FIRES is centered at $(10^h56^m58.26^s,
    -03^d37^m0.53^s)$. Each CDFS pointing is limited by the $8.2$
    arcminute diameter field of view (FOV) of the LDSS-3 spectrograph,
    shown as the thick black circle. The FIRES area is a 5.5 x 5.3
    arcminute region which falls completely within the LDSS-3 FOV. All
    of the galaxies targeted are shown as open circles while those
    galaxies with observed emission lines (not necessarily [OII]) are
    overlaid with filled circles.}
  \label{fig:ROLES_KG650_pointings}
\end{figure*}

ROLES demonstrated that the sSFR-mass relation evolves steadily with redshift, in a nearly mass-independent way, so the SFR density (SFRD) evolution is characterised primarily by an evolution in normalization only \citep{Gilbank2010_ROLESII}.  However, there is a hint that the low-mass end of the sSFR-mass relation becomes steeper at $z\sim 1$ \citep{Gilbank2011}, suggesting that the lowest-mass galaxies formed their stars later, and on longer timescales.   Surprisingly, despite the small fields covered, \citet{Li2011} found a clear environmental dependence amongst the star--forming population.  Star--forming galaxies in only moderately (factor $\sim 15$) overdense regions at $z=1$ appear to have {\it higher} SFR, a result that is opposite to the (weak) trend seen locally.  This is qualitatively consistent with results  from some other surveys 
\citep[e.g.][]{Elbaz2007,I+09,Sobral2011}.

Here, we adopt the ROLES methodology \citep{Gilbank2010_ROLESII} to
explore star formation and its environmental dependence amongst low-mass galaxies over the redshift range $0.62 < z < 0.885$.  This provides an intermediate link between ROLES at $z=1$ and the local Universe, using consistent galaxy selection and SFR measurement methods.  Moreover, the redshift range and fields were chosen to include highly overdense regions, including the well-studied MS1054-03 galaxy cluster \citep[e.g.][]{vD+00,Forster2006}.  Thus the data span a wider range in environment compared with the ROLES data.

This paper is presented as follows. \S\ref{sec:1_method} describes the
survey and image reduction methodology, while details of the emission
line detection procedure are presented in 
\S\ref{sec:2_automated_line_finding}.  The basic measurements,
corrections, and limiting values are presented in
\S\ref{sec:3_science_analysis}.  Our results are shown in
\S\ref{sec:4_results}, and we compare our results on the environmental
independence of sSFR with published results at $z=0$ and $z=1$ in
\S\ref{sec:5_discussion}.  Finally, our conclusions are summarized in 
\S\ref{sec:6_conclusions}. AB magnitudes
are used throughout unless otherwise stated and we use a $\Lambda$CDM
cosmology of $H_0 = 70~\mbox{km}~\mbox{s}^{-1}~\mbox{ Mpc}^{-1}$, $\Omega_M = 0.3$, and
$\Omega_{\Lambda} = 0.7$. Finally, note that all ROLES SFRs have been corrected
using the empirical stellar mass dependent relationship determined in
\citet{Gilbank2010}, and described in \S\ref{sec:3_science_analysis}.  

\section{Data Acquisition \& Reduction}
\label{sec:1_method}
The design and implementation of the present survey
is similar to our previous work at
$z=1$  \citep[hereafter referred to as
ROLES]{Gilbank2010_ROLESII}. In this section we review the target
selection criteria, observation strategy, and image reduction steps. 

\subsection{Target Selection}
\label{sec:target_selection}
Targets were selected based upon their $K$-band magnitudes, $22.5 < K <
24$, and their photometric redshifts as provided by
\citet[FIRES]{Forster2006} and Mobasher \& Dahlen
\citeyearpar[CDFS]{Mobasher2009}. During the initial survey mask design
phase, photometric redshifts were used to prioritize those targets
which were expected to lie within our redshift range of $0.62 < z <
0.885$, considering the probability distribution of the photometric redshift.  Galaxies with large photometric redshift uncertainties, or which were expected to lie outside our target redshift range, were also included in the mask design, with lower priority.  As with ROLES, the high sampling density in these fields means that the details of the prioritisation play a limited role, and the photometric redshift preselection does not constitute a strong prior.

Our survey consists of two pointings in the Great Observatories Origins
Deep Survey (GOODS) region of the Chandra Deep Field South \citep[CDFS,
e.g.][]{Wuyts2008} and one pointing in the MS1054-03 cluster region of
the Faint Infra-Red Extragalactic Survey \citep[FIRES,
e.g.][]{Forster2006,Crawford2011}
field, for a total of 18 masks.
The three pointings are shown in Figure \ref{fig:ROLES_KG650_pointings}
with observational targets and emission line detections (described
below) indicated.

\subsection{Observations}
\label{sec:observations}
All spectroscopic observations were obtained using the 6.5 meter
Magellan (Clay) telescope. Multi-object spectroscopy for our 1946
targets was provided by the Low Dispersion Survey Spectrograph 3
(LDSS-3). The spectra were dispersed by the \textit{medium red} grism
(300 lines/mm) which has a dispersion of approximately
$2.65\mbox{\AA/pixel}$ at $6500\mbox{\AA}$ and a relatively uniform
throughput across the KG650 wavelength range. Combined with the plate
scale of $0.189''\mbox{/pixel}$ and survey mask slit width of 0.8'',
the resolution is $11.2\mbox{\AA}$ FWHM.  

\begin{figure}
  \includegraphics[width=1.0\linewidth]{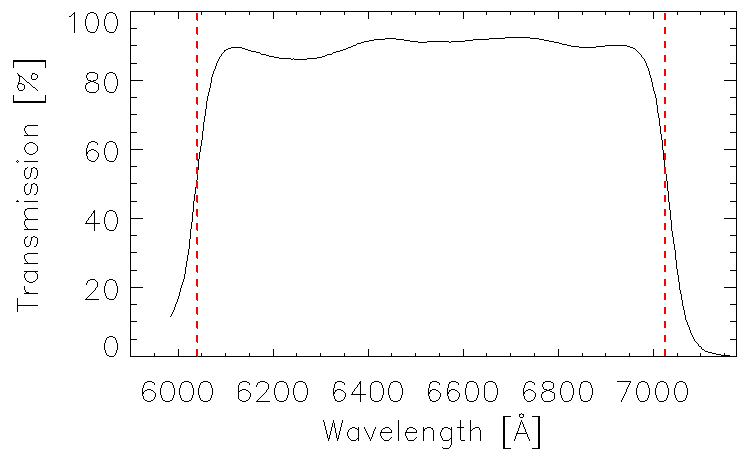}
  \caption{The transmission curve for the custom designed KG650
    filter. The vertical dashed lines at
    $\lambda~=~6040,7025\mbox{\AA}$ indicate the spectral range
    considered by our survey, and correspond to the full width at half
    maximum (FWHM) locations of the transmission curve. Since the
    survey specifically targets [OII] emission at $3727$\AA, the survey redshift range is also
    restricted by the filter FWHM and is taken to be
    $0.62~<~z~<~0.885$. }
  \label{fig:KG650_trans_curve}
\end{figure}

The spectral wavelength range was limited to approximately
$650~\pm~50~\mbox{nm}$ by a filter, herein referred to as KG650. The
transmission curve for the KG650 filter is shown in Figure
\ref{fig:KG650_trans_curve}. From this transmission we define our
sensitivity range as 
$6040\mbox{\AA}~\le~\lambda_{obs}~<~7025\mbox{\AA}$.

The design of the survey masks was driven by the Nod \& Shuffle
\citep[N\&S, ][]{Glazebrook2001,Gilbank2010_ROLESII} observing strategy.  The principle advantage of this technique is
that it allows for accurate sky subtraction at red wavelengths, where
the sky brightness is dominated by rapidly varying emission lines.  Target
slits were 0.8\arcsec\ wide by 3.0\arcsec\ long, which allowed for nearly 200
objects to be targeted per mask given the LDSS-3 FOV of 8.2\arcmin\ diameter,
and the requirement that dispersed object spectra must not
overlap. We used a N\&S cycle time of 60 seconds and a slit nod
distance of 1.2\arcsec.

The exposure time for each mask is shown in Table
\ref{tab:ROLES_masks}. From the 18 masks combined, a total of 3292
objects were targeted. This includes objects which were
targeted in multiple masks, and filler objects that do not satisfy our
primary selection criteria.  The total number of primary targets was 2770 (including
duplicates), of which 1946 are unique.  These are split between 640 in
FIRES and 1306 in CDFS. There were cases where slits were cut too close
to mask edges, or to each other, and thus yielded unusable spectra. The
number of bad slits per mask typically ranged between 2 and 7.
CDFS masks 40 and 41 had an unusually large number of bad slits,
with 13 and 14 respectively. For the FIRES field, three exposures per
mask were acquired while for CDFS we adopted a shorter exposure time
per mask and acquired four frames per mask. The total exposure time per
mask was identical for FIRES and CDFS except for CDFS masks 40 and 41
where we acquired an extra 30 minute exposure for each due to time
remaining in the observing programme. 

\begin{table}
  \begin{center}
  \caption{Spectroscopic masks with their corresponding target counts,
    total exposure times, and typical guide star seeing conditions. The
    two pointings in CDFS are labeled as CDFS.1 and CDFS.2. The total
    number of unique survey targets is 1946. The number of targets,
    $N_{\rm targets}$, refers to the number of usable main survey spectra,
    excluding filler objects but including duplicates.
} 
    \begin{tabular}{ c | c | c | c | c}
      Mask ID&Field&$N_{\rm targets}$&Exposure Time&Seeing\\
      && &(hrs)&(arcsec)\\
      \hline
      mask24&FIRES&115&2&0.5\\
      mask25&FIRES&112&2&0.6\\
      mask26&FIRES&100&2&0.63\\
      mask27&FIRES&95&2&0.53\\
      mask28&FIRES&99&2&0.6\\
      mask29&FIRES&105&2&0.72\\
      mask30&CDFS.1&191&2&1.0\\
      mask31&CDFS.1&180&2&0.79\\
      mask32&CDFS.1&178&2&0.8\\
      mask33&CDFS.1&175&2&1.0\\
      mask34&CDFS.1&178&2&1.0\\
      mask35&CDFS.1&171&2&0.66\\
      mask36&CDFS.2&167&2&0.86\\
      mask37&CDFS.2&167&2&0.92\\
      mask38&CDFS.2&164&2&0.88\\
      mask39&CDFS.2&160&2&1.19\\
      mask40&CDFS.2&156&2.5&0.82\\
      mask41&CDFS.2&154&2.5&0.74\\
    \end{tabular}
    \label{tab:ROLES_masks}
  \end{center}  
\end{table}

\subsection{Image Reduction}
\label{sec:image_reduction}
The FITS image files created by LDSS-3 were processed through an image
reduction pipeline similar to that described in the Carnegie
Observatories COSMOS (Carnegie Observatories System for MultiObject
Spectroscopy) Cookbook\footnote{See
  http://obs.carnegiescience.edu/Code/cosmos/Cookbook.html}, with
custom-written IDL routines to supplement the existing software when required.

\subsubsection{Initial Frame Combining}
\label{sec:init_frame_combine}
LDSS-3 was read out in two amplifier mode meaning that each mask
exposure consisted of two raw FITS images. These images were combined
using the COSMOS ``stitch'' routine with gain parameters set according
to each specific amplifier and dewar parameter set to LDSS3-2.
The ``stitch'' routine removes bias and corrects for differences between amplifier gains in LDSS-3 so no further bias removal was necessary after this stage.

\subsubsection{Bad Pixel Mask from Charge Traps \& Cosmic Rays}
\label{sec:ns_dark_bpm}
A bad pixel mask (BPM) was created using Nod \& Shuffle dark frames
(N\&S darks) acquired 
using the same N\&S parameters as for 
the science frames. The resulting image is one which is mostly
dark (predominantly read noise counts) with streaks of bright
pixels indicating bad pixels created by charge being trapped by
individual pixels in the CCD array. The streaks correspond to the
charge shuffle distance and direction defined by the N\&S strategy. A
BPM was made from a N\&S dark by dividing the original N\&S dark frame
by a 1x3 boxcar smoothed version (smoothing done in the direction
perpendicular to the shuffle direction) of the same frame. Bad
pixels appeared brighter in the ratio of the frames and were recorded
as being bad in the BPM. Since several N\&S darks had been acquired,
they were each processed in the same manner and finally median combined
into one single BPM. 

As many masks were observed with only three exposures, we used the IRAF
{\sc cosmicrays} task to identify the locations of cosmic rays (CR) in
each frame based upon user-defined threshold levels and cosmic ray
shapes. The pixel locations were recorded as a unique BPM for each
exposure, and this resulted in superior quality image stacks during
median combination.

\subsubsection{Wavelength Calibration}
\label{sec:wavelength_calibration}

The COSMOS {\sc apertures} routine was used to make predictions of initial
positions of the slit centers in each mask. These positions were
compared to the actual slit center positions as imaged through the
optical path and corrected (to less than 1 pixel difference) using the
{\sc align-mask} routine. The positions of known arc lines were predicted
for the arc calibration frames using the {\sc align-mask},
{\sc map-spectra}, and {\sc spectral-map} routines.  
The COSMOS {\sc adjust-map} routine was adequate for providing an
initial wavelength calibration solution for most slits in a
given mask. However there remained several cases where analysis of sky emission lines
revealed inaccurate calibration.  For this reason an IDL routine was used to determine a third
order wavelength calibration solution to all of the slits in each
mask, based on the position of these emission lines. Final emission
line position residuals were typically $\lesssim 
0.7\mbox{\AA}$.

\subsection{Creation of Stacked Frames}
\label{sec:creation_of_stacked_frames}
For most masks the individual exposures to be stacked were
acquired on different dates. As masks were interchanged in the optical
path frequently and the telescope was at different orientations while
tracking the target field at different times of the year, differences
in mask flexure, rotation, and shifts were introduced
between one exposure and another. 
The transformations between science frames and an arbitrary reference
frame were determined
based on the common sky emission line centroid positions in each
frame. The transformations commonly required a small rotation, shifts
in the $X_{CCD}$ and $Y_{CCD}$ directions, and on occasion a
multiplicative scaling. The IRAF task {\sc geomap} computed these transformations while
the {\sc geotran} task was used to apply them to each non-reference
frame to be stacked. 

The applied {\sc geomap/geotran} transformations accounted for
differences in slit positions from one exposure frame to another.
However, there were also cases where the target galaxy within a slit
varied slightly in position between the frames to be stacked. To
rectify this, another IDL program was written to determine and apply any further
required shifts in the spectral and spatial directions based upon a
list of bright emission lines identified by eye and found in each frame
to be stacked. 
Any such shifts were
typically $\sim1~\mbox{pixel}$ in the $X_{CCD}$ and/or $Y_{CCD}$
directions.

The steps required to create the stacked signal frame from the N\&S
observations are described below.  Individual exposure frames are
labeled as \textit{A} and \textit{B}, and the recipe can be extended
to an arbitrary number of frames.
\begin{enumerate}
        \item Shift frame \textit{A} by 16 pixels in the spatial (``y'') direction, to get a new frame \textit{$A_1$};
        \item Perform the subtraction \textit{A - $A_1$} to get a new
          frame \textit{$A_2$}.  This is the \textit{sky subtracted} frame;
        \item Shift frame \textit{$A_2$} by 6 pixels in the spatial (``y'') direction, to get a new frame \textit{$A_3$};
        \item Perform the subtraction \textit{$A_2$ - $A_3$} to get a new frame \textit{$A_4$}. This frame is the ``positive'' image frame;
        \item Repeat steps (i) through (iv) for all individual exposure frames;
        \item Determine if individual frame flux scaling is necessary for each frame based upon the flux level ratios of several manually identified emission lines common between the brightest frame and the frame being scaled;
        \item Apply further (small) frame shifting if necessary based upon the centroided positions of the identified emission lines used in step 6;
        \item Median add the ``positive'' image frames, \textit{$A_4$} + \textit{$B_4$} = \textit{C}.
\end{enumerate}
The IRAF task {\sc imcombine} was used to median stack the individual
``positive'' image frames. 
The bad pixel mask described in \S\ref{sec:ns_dark_bpm}, which includes
identified cosmic rays, was used to ignore pixels during the
combination.

A corresponding noise frame
was created in a manner similar to the stacked signal frame, as
described 
by \citet{Gilbank2010_ROLESII}.  In the equations below the subscripts \textit{ij} refer to the $ij^{th}$ pixel of the frame. 
\begin{enumerate}
        \item Apply the same frame flux scaling, determined in step (vi) of the stacked science frame creation recipe, to each \textit{sky added} frame;
        \item Apply the same (small) frame shifts, determined from the locations of common bright emission lines used in step (vii) of the stacked science frame creation recipe, to each \textit{sky added} frame;
        \item Stack (median add) the \textit{sky added} image frames to get a new frame, $|<sky>|$;
        \item Add in the LDSS-3 read noise, \textit{R}. The read noise must be added in twice since the median combined frame consists of a shifted frame added to a non-shifted frame, each containing read noise. The read noise adjusted frame is calculated as follows: 
        \begin{equation}
                N_{indiv,ij} = \sqrt{(\sqrt{|<sky>|_{ij}})^2 + 2(R^2)}
        \end{equation}
        \item Scale frame $N_{indiv,ij}$ by the number of individual science frames used in the median combination, $n_{frames}$,
        \begin{equation}
                N_{com,ij} = \frac{N_{indiv,ij}}{\sqrt{n_{frames}}}
        \end{equation}
        \item Shift the frame $N_{com,ij}$ by 6 pixels in the spatial (``y'') direction, to get a new frame $N_{com,ij}'$;
        \item Perform the quadrature addition of these last two frames to get the final noise frame:
        \begin{equation}
                N_{ij} = \sqrt{(N_{com,ij})^2 + (N_{com,ij}')^2}
        \end{equation}
\end{enumerate}

\section{Emission Line Detection}
\label{sec:2_automated_line_finding}
\subsection{Creation of Signal-to-Noise Frame}
\label{sec:creation_of_s2n_frame}
To identify faint emission lines, a normalized 2-D convolution kernel, $k_{em}$, was created which had the same Gaussian shape as a typical bright emission line (FWHM=[5.5,3.5] pixels), and was convolved with the signal ($S$) and noise ($N$) frames to give flux-conserved, convolved signal and noise frames, according to:

\begin{equation}
        S_{conv,ij} = S_{ij} \otimes k_{em}
        \label{eq:conv_signal}
\end{equation}
and
\begin{equation}
        N_{conv, ij} = \sqrt{N^2_{ij} \otimes k^2_{em}}.
        \label{eq:conv_noise}
\end{equation}

The next step was to estimate the continuum found in the original
signal frame. Similar to the convolution of the signal and noise
frames, a convolution was again performed on the raw signal frame,
using a 2-D normalized averaging kernel, $k_{cont}$:
\begin{equation}
        C_{ij} = S_{ij} \otimes k_{cont}.
        \label{eq:contnm}
\end{equation}
The shape of the kernel 
consisted of a \textit{zero}
central region (20 pixels spectral by 3 pixels spatial) and two
sidebands (also each 20 pixels spectral and 3 pixels spatial). The
sidebands had the same Gaussian FWHM of 3.5 pixels in the spatial
direction as the emission line kernel for their entire spectral length
of 20 pixels. Convolving the kernel with the raw science frame provided
an estimation of the continuum for the pixel located at the center of
the kernel. The zero region was
included so that the continuum estimate was not biased by the presence
of an emission line.  This provides a continuum estimate that is
effectively 
an average of the flux in the spectral and spatial
directions, in the ``wings'' of the pixel for which the continuum was
being determined.

The noise due to the continuum, $N_{cont, ij}$, was calculated by convolving the emission line kernel, $k_{em}$, with the estimation of the continuum frame, $C$, as follows:

\begin{equation}
        N_{cont, ij} = \sqrt{C_{ij} \otimes k^2_{em}}
        \label{eq:contnm_noise}
\end{equation} 
and the total noise frame, $N_{total}$, was calculated by adding in quadrature the convolved raw noise frame with the convolved continuum noise estimate
\begin{equation}
        N_{total, ij} = \sqrt{N^2_{ij} + N^2_{cont,ij}}
        \label{eq:total_noise}
\end{equation} 
In most cases the noise is dominated by sky line emission, but the
continuum noise is not entirely negligible.

This procedure accurately accounts for statistical noise in our
spectra, but may not account for low-level systematics resulting from
weak charge traps, sky emission residuals (minimized but not entirely
eliminated with N\&S cycles of 60s), or overlapping spectra.
We analyzed the {\it rms} fluctuations in the final science frames and
compared this with the associated noise estimate, for each mask.
Specifically, thirty equally spaced
``test'' locations were chosen along the center line (distributed in
the spectral direction) of each slit in a given mask, for both the
stacked science and stacked noise frames. The mean pixel value for each
slit test location in the science frame was determined by taking the
mean of the pixel values within two 60 pixel sidebands, located to
either side of the test location. The fluctuation of the test location
pixel value from the mean was then simply the actual pixel value
subtract the mean value. For every slit in the mask, the science
fluctuation 
($\sigma_S$) and noise
value ($\mu_n$) for thirty test locations were recorded. Histograms of the science
fluctuations and corresponding noise values were then fitted with
Gaussians. Finally, the ratio of the best-fit Gaussian standard
deviation of the science frame fluctuations and the Gaussian mean of
the noise values gave the ``noise correction factor'' (NCF):
\begin{equation}
        NCF = \frac{\sigma_{S}}{\mu_{n}}
        \label{eq:NCF}
\end{equation}
A typical noise correction factor was $\sim$1.2, indicating that
residual systematics amount to an additional 20\% on top of the
statistical noise.  

The final \snframe~was calculated as 
\begin{equation}
        \left\lbrace \frac{S}{N} \right\rbrace_{ij} = \frac{S_{conv,ij} - C_{ij}}{N_{total,ij} \cdot NCF}
        \label{eq:s2n}
\end{equation}

\subsection{Emission Line Finding}
\label{sec:EL_finding}
The central five rows of each spectrum was extracted from the 2D frame,
to minimize effects near slit edges that affect line detection.
For every pixel above a 
\snthresh\ of $3$, an ``n-connected neighbour'' search was performed to locate
connected neighbouring pixels that also exceed this threshold.
A candidate detection then consists of two or more connected pixels;
if multiple detections were separated by five pixels or
less, they are combined into a single detection.
Detections found within three pixels of the spectral ends of the
extracted spectrum, and those that were due to overlaps of $0^{th}$ order spectra,
were excluded.
The resulting list was visually inspected, and obvious false detections
(due in general to overlapping spectra or missed cosmic rays) were
manually removed.

\subsubsection{Catalogue Purity}
\label{sec:emline_cat_testing}
\begin{figure}
        \includegraphics[width=1.0\linewidth]{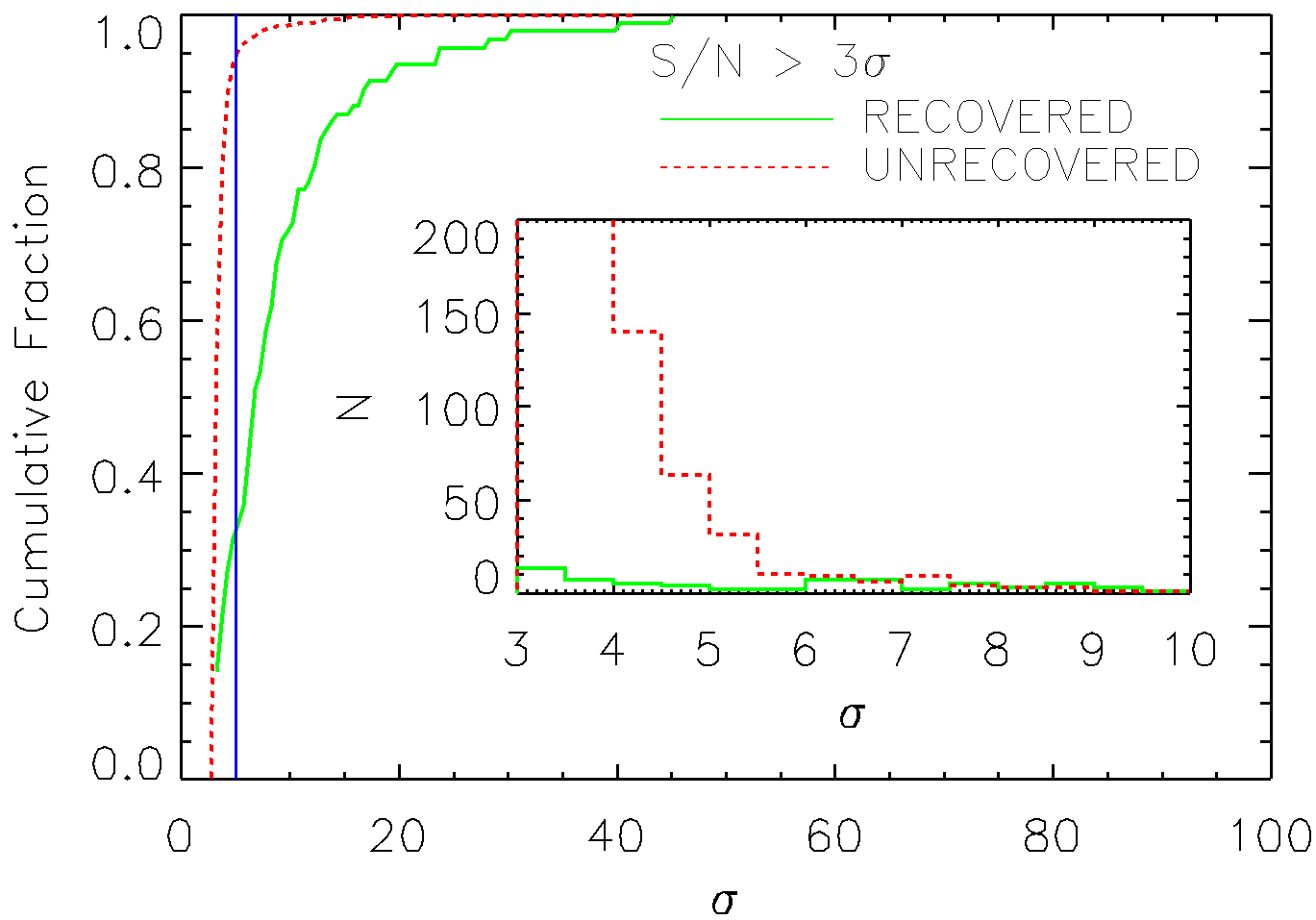}
        \caption{{\it Main Panel:} The cumulative fraction of the number of
          recovered (solid green curve) vs. spurious (dashed red curve)
          emission lines for galaxies which were targeted in multiple
          masks, as a function of significance. {\it Inset: } Histogram of
          the number of recovered and spurious emission lines. This demonstrates that 95\% of spurious
          detections occur below $5\sigma$ (highlighted as the solid
          vertical blue line).} 
        \label{fig:duplicate_target_det_recovery_vs_spurious}  
\end{figure}\begin{figure}
        \includegraphics[width=1.0\linewidth]{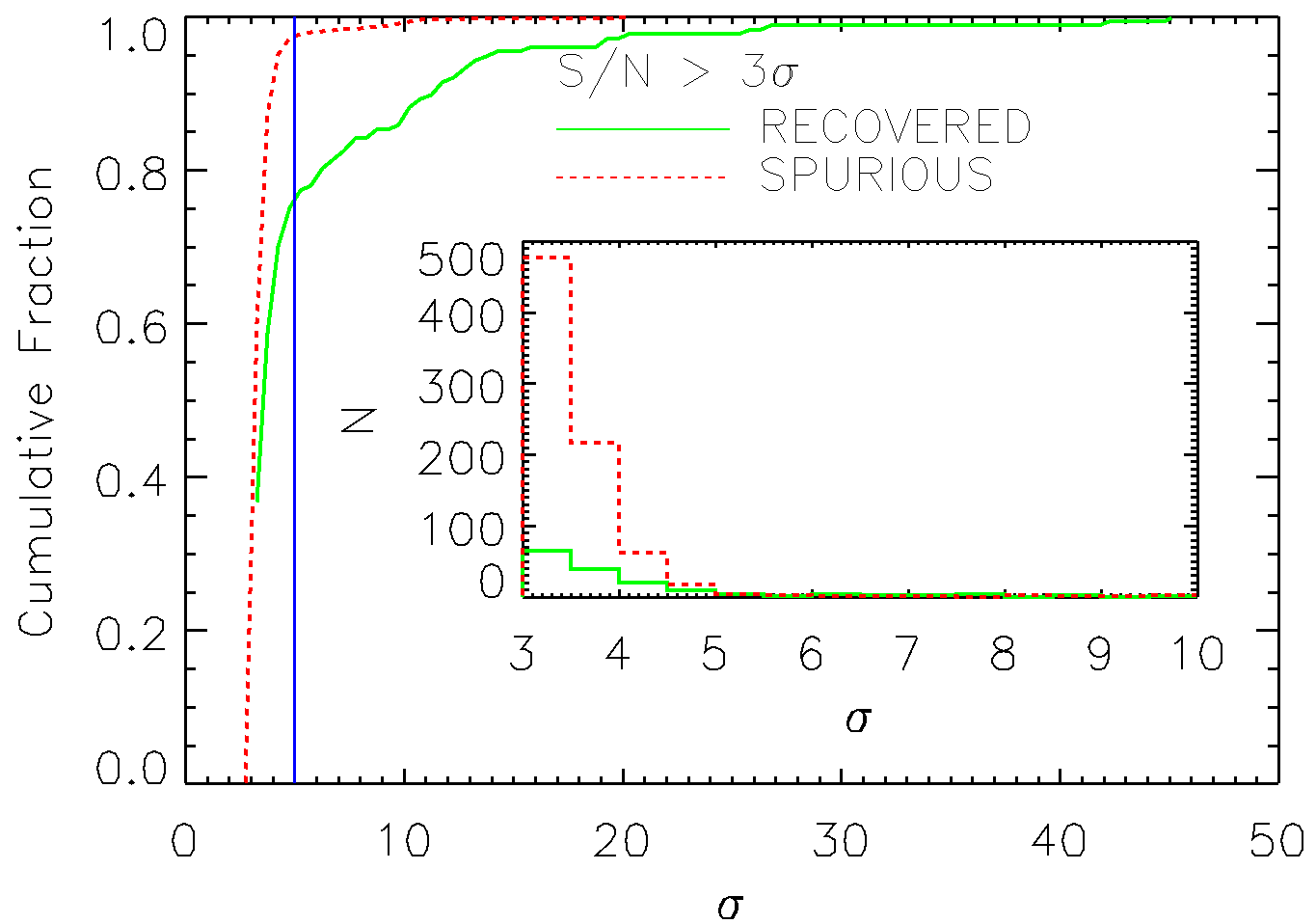}
        \caption{{\it Main Panel:} The cumulative fraction of the number of
          recovered (solid green curve) vs. spurious (dashed red curve)
          emission lines for galaxies which contain at least two
          emission lines and lead to self-consistent redshifts, as a
          function of significance. {\it Inset:} Histogram of the number of
          recovered and spurious emission lines providing
          self-consistent redshifts for the galaxies containing
          them. It is clear from the cumulative fractions that
          $\sim97\%$ of spurious detections occur below $5\sigma$
          (highlighted as the solid vertical blue line).} 
        \label{fig:plasible_line_consistent_redshift_recovery_vs_spurious}
\end{figure}
The $3\sigma~(S/N\ge3)$ catalogue was internally tested in two ways to
determine the purity, following \citet{Gilbank2010_ROLESII}.
We first consider the reproducibility of
emission lines for the 412 galaxies that were targeted on more than one mask. 
For these galaxies, detection lists were compared and emission lines were considered
to match if their wavelengths were within $\pm6.5\mbox{\AA}~
(2.5~\mbox{pixels})$ of each other. If a detection was found in all
of the masks the galaxy was targeted in, then it was considered fully
recovered; otherwise it was considered spurious. This is therefore a
conservative estimate of the purity.  
The results of this test are shown in Figure
\ref{fig:duplicate_target_det_recovery_vs_spurious} where it is clear
that 95\% of spurious detections occurred in detections below $5\sigma$.

An independent test of purity is to consider spectra for which more
than one candidate is detected.  The wavelengths of these candidates
were compared with expected sets of lines, which are likely to appear
in only a small number of combinations:
the \textit{$H_{\beta}-[OIII]$} complex ($H_{\beta}, [OIII]_{4959},
[OIII]_{5007}$), and the $[NeIII]_{3869},
[OII]_{3727}$ pair.   
Candidates were considered real detections if their line ratios
matched one of these combinations.
In the case that the lines did not correspond to an expected set, the
line with the highest significance was considered real, while the
next-highest significance line was considered spurious; lower
significance lines were omitted for the purpose of this test.
Figure
\ref{fig:plasible_line_consistent_redshift_recovery_vs_spurious} shows
the S/N distribution of these real and spurious lines;
$\sim97\%$ of spurious lines have significance less than $5\sigma$.

From these tests, we conclude that $>95$\% of false detections occur
below a 
significance threshold of $S/N\ge5\sigma$; thus we only consider
detections above this limit in our analysis.

\subsection{Emission Line Flux Determination}
\label{standard_star_calibration}
Flux calibration was based on the spectrophotometric standard star HD
49798 \citep{Bohlin1992}.
Emission line fluxes and their errors were measured from the
stacked raw science frames with the continuum estimation removed, and
the stacked raw noise frames. The bad pixel masks were incorporated to
eliminate bad pixels and cosmic rays.  
For each detection, the
centroid position of the flux was found within a 15 x 17 pixel box,
initially centered on the location of the highest significance pixel in
the emission line. The total emission line flux was taken to be the sum
of flux within a 7 x 5 pixel region about this centroid.
The line flux error was calculated for the same pixels, based on the
noise spectrum.

To account for varying photometric conditions, we compare the flux in the continuum measured from
the spectra on each mask with photometric data from public catalogues.
For the CDFS field we use the R-band magnitudes from FIREWORKS
\citep{Wuyts2008}, which covers almost exactly the same wavelength
range as our spectroscopy.
For the FIRES field, the available photometry does not include
R-magnitudes, so we interpolated between the HST WFPC2 F606W and F814W
filters.  For each mask we calculate the average offset between the
flux in our spectra and the continuum flux measured from the imaging.
We use this to identify the most photometric mask in each field, and
the offsets from this mask for all of the others.  We then correct the
flux calibration for the non-photometric masks, to match this reference
frame.  The correction is typically $\sim 0.5$ mag, with a maximum of 1
mag.  

\begin{figure}
        \includegraphics[width=1.0\linewidth]{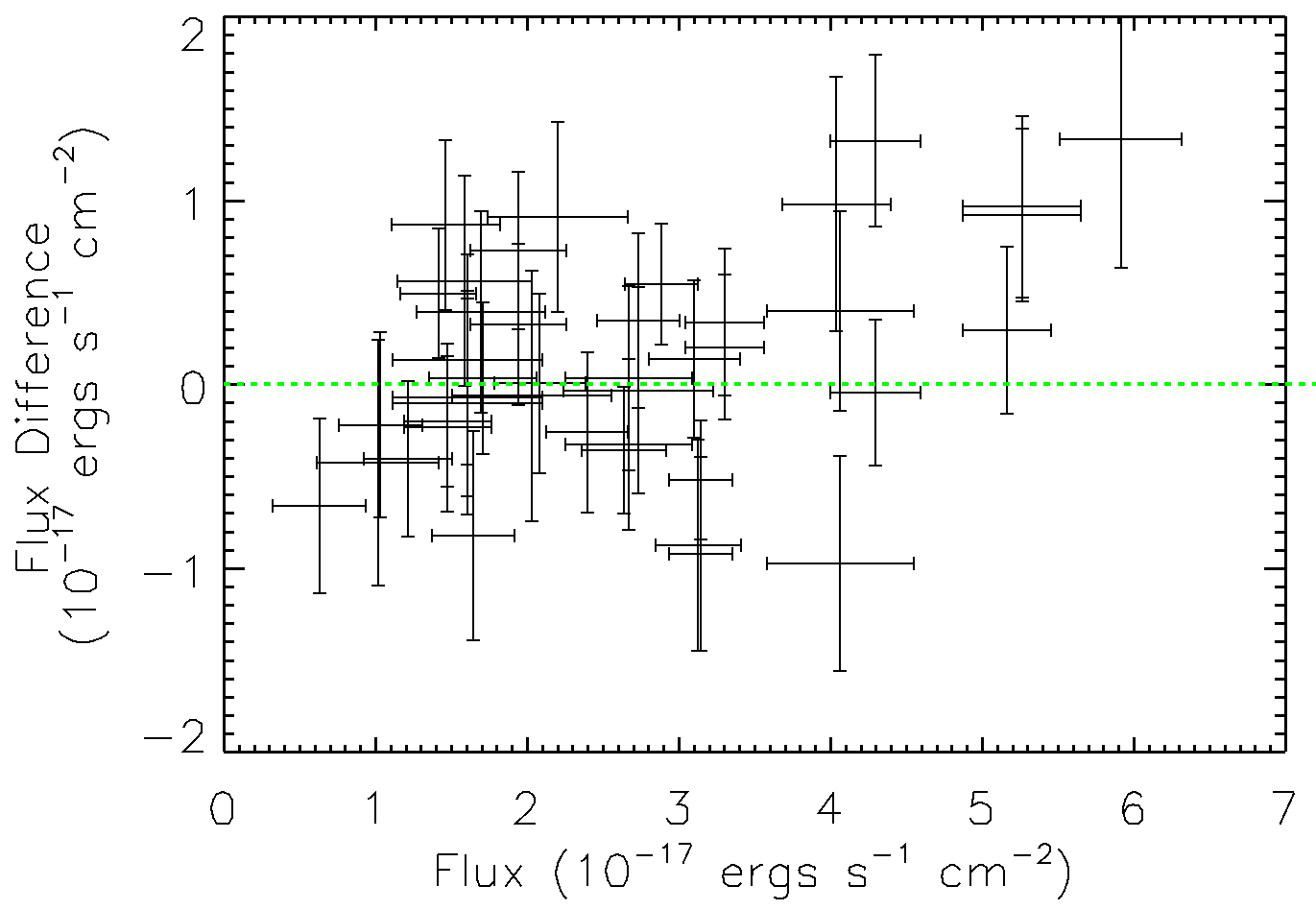}
        \caption{ The difference in $\ge5\sigma$~emission
          line fluxes for matching detections found in multiply-imaged
          galaxies as a function of line flux. The zero difference line
          is shown as the horizontal green dashed line. } 
        \label{fig:roles_R_band_mask_to_mask_flux_error_analysis}
\end{figure}

We take advantage of galaxies within our $5\sigma$~linelist that
were imaged on multiple masks, to further check the consistency of the
flux calibration and our uncertainty estimates.  
The flux differences for separate observations 
were determined and
plotted as a function of line flux, as shown in Figure
\ref{fig:roles_R_band_mask_to_mask_flux_error_analysis}. As expected,
the matching line flux differences are scattered about zero.  The
significance of each difference is obtained by dividing by the flux
uncertainties added in quadrature.  The significance distribution has a
standard deviation of $\sigma\sim1.28$, but a
Kolmogorov--Smirnov test cannot distinguish between this distribution
and a normal distribution with $\sigma=1$.  
Thus we conclude that uncertainties in relative flux calibration from
mask-to-mask are negligible.

We looked for a correlation between the flux difference compared with
the photometry and galaxy size.  While there does appear to be a
correlation in the expected sense, that larger galaxies are missing
more flux in our spectroscopy, there is a lot of scatter from object to
object.  We have elected therefore not to apply an aperture correction,
but note that the fluxes for our largest galaxies are likely underestimated.
Only three of our spectra overlap with spectra obtained by
\citet{Vanzella2008}; while this comparison shows our flux calibration
is consistent with theirs, there are not enough objects in common to
state this with a high degree of confidence.  Thus we expect our line
fluxes are dominated by this systematic uncertainty in flux
calibration, which is likely at least a factor of $\sim 2$ with a
dependence on galaxy size.

\begin{figure}
        \includegraphics[width=1.0\linewidth]{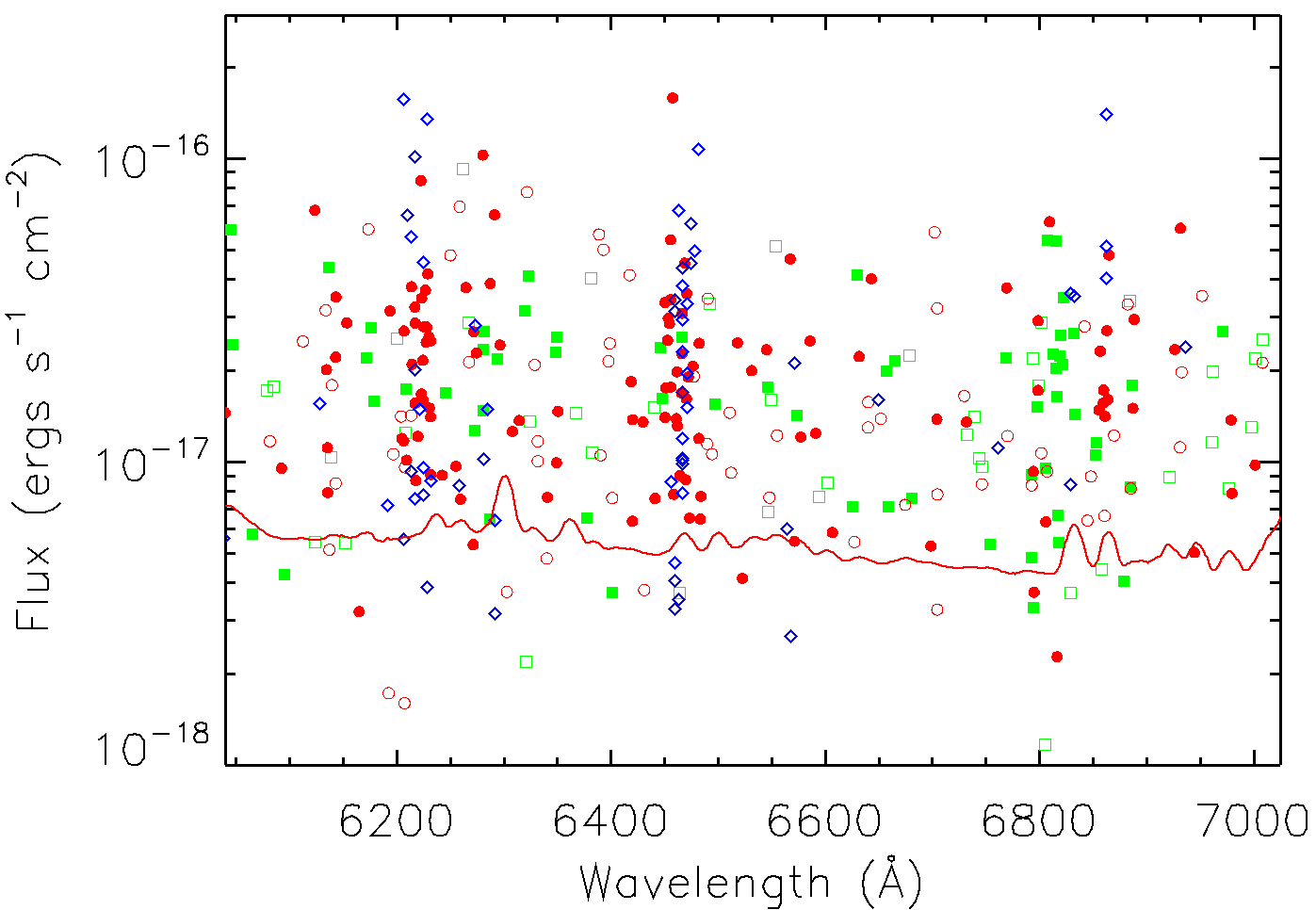}
        \caption{Fluxes for emission lines detected with $5\sigma$ significance are shown as
          a function of wavelength.  
Detections are
         divided into the CDFS ROLES (solid/open red circles), FORS2
          (blue diamonds), and FIRES (solid/open green squares)
          fields. The average $5N_{ij}$ noise flux limit of all
          masks is overplotted as a solid line. 
}
        \label{fig:noise_flux_culled_dets}
\end{figure}

The wavelength-dependent flux limit was determined for each survey mask from the
associated noise propagated through the analysis pipeline.  
For each mask, 
the average noise spectrum $\sigma_\lambda$ was determined from all the dispersed
spectra in the mask, and secure detections were then defined as those
brighter than $5\sigma_\lambda$.  
Figure \ref{fig:noise_flux_culled_dets} shows the flux of all detected lines, and the average 5$\sigma_\lambda$
noise level for all masks.  Unless otherwise stated, analysis in this paper excludes lines that fall below the average flux limit shown here,
to enforce a uniform limit.

\subsection{Line Identification}
\label{sec:line_id}
For galaxies in the $\ge5\sigma$~catalogue where more than one emission
line was detected with an appropriate wavelength separation,
identification was straightforward.  However, there are only
41 such candidates, 15 of which are identified as [OII].
The remainder of the catalogue consists of single emission
lines, for which we rely on  
the photometric redshift
probability distribution functions (PDFs) to identify the line, as
described in \citet{Gilbank2010_ROLESII}.
The relative likelihood of a line being [OII] was assigned to each detection in
our $5\sigma$ catalogue by determining the ratio of the probability of
the emission line being [OII] to the total probability of being either
$[\mbox{MgII}]$, $[\mbox{CIV}]$ or one of the
$H\beta-[OIII]$ complex.   The probabilities were calculated by
integrating the photometric redshift PDFs over the redshift ranges
corresponding to the lines of interest in our spectral window. 

\begin{figure}
        \includegraphics[width=1.0\linewidth]{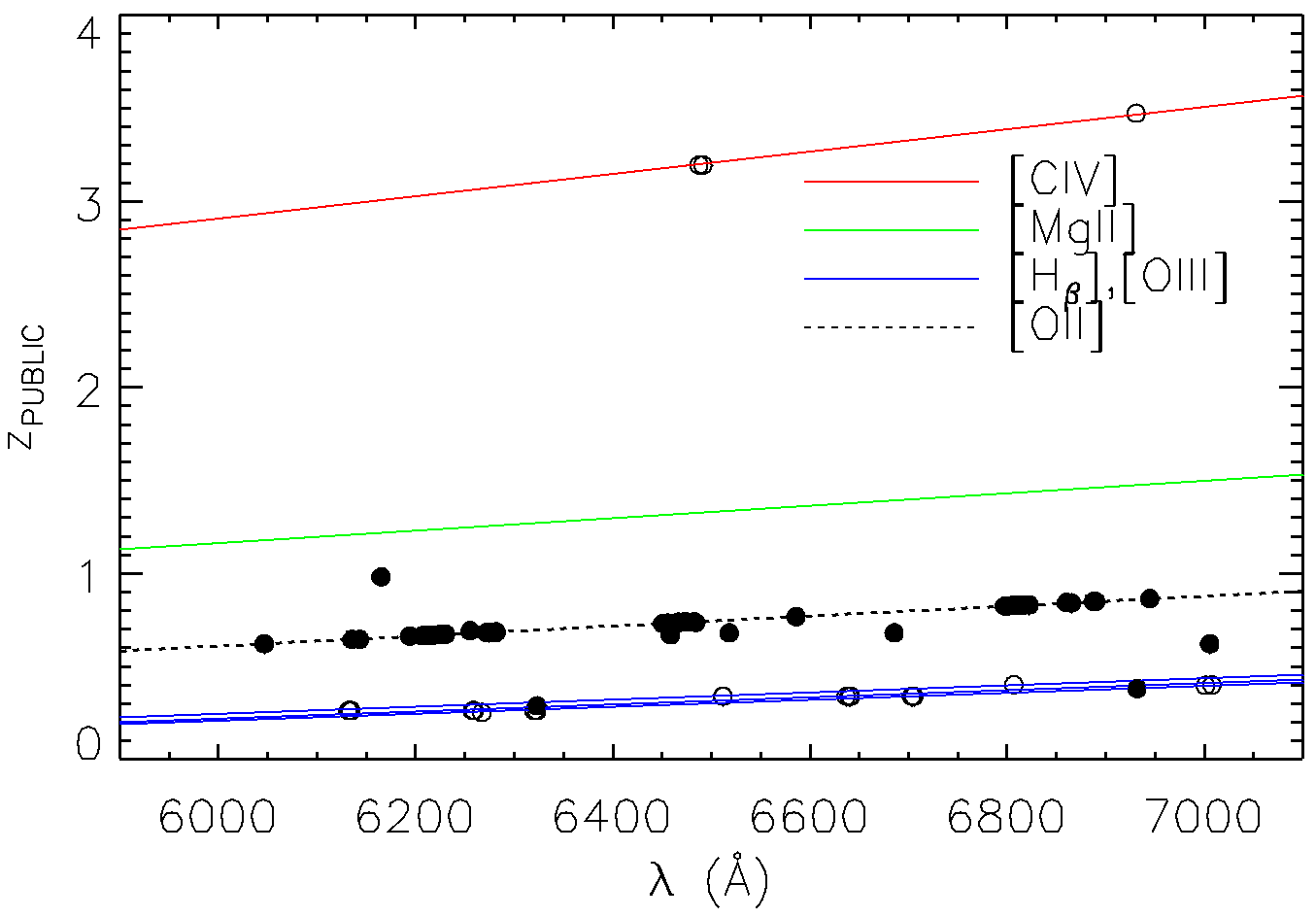}
        \caption{The published redshift, where available, is shown for galaxies with candidate
          emission lines in our catalogue as a function of the line wavelength. Angled lines
          show the relation expected for seven different emission lines as
          indicated; in particular 
the black
          dashed line indicates our target line of [OII].  Filled circles correspond to detections with
          $\mathcal{L}[OII]>0.9$.
} 
        \label{fig:roles_public_zsp_id}
\end{figure}

For a few lines, publicly available spectroscopy covering a larger
wavelength range are available; nine from \citet[][MS1054-03]{Crawford2011}
and 73 from \citet[][FIREWORKS]{Wuyts2008} have secure redshifts.
Figure
\ref{fig:roles_public_zsp_id} shows these secure redshifts as a
function of the wavelength of the detected emission line in our
sample.  Straight lines indicate
the redshift--wavelength relation for different emission lines,
including [OII] (black dashed line). 
The detections denoted by solid symbols were those which had
an integrated photometric redshift probability of being [OII]. 
Most of the emission lines which had a high probability of being [OII]
($\mathcal{L}[OII]>0.9$) are confirmed as such from the public
redshift.  
Of the 82 matched detections shown in the figure only three appear
inconsistent with their published redshift, suggesting that the purity
of our sample is over 95 per cent.

\subsection{Final $5\sigma$ Catalogue}
\label{sec:final_cat}
The final catalogue contains all detections with $S/N \ge 5$, and a
redshift determined following the line identification procedure
described above.
The detections were also given a \textit{line quality} flag as follows:
\begin{enumerate}
        \item[0:] Line is not likely [OII] based on the photometric
          redshift PDF, and there is no existing public redshift
        \item[1:] Photometric redshift is consistent with the detection being [OII]
        \item[2:] Photometric redshift is consistent with the detection
          being [OII], confirmed by detection of [NeIII].  
        \item[3:] Photometric redshift is consistent with the detection
          being [OII], confirmed by a published redshift.
\end{enumerate}

\section{Analysis}
\label{sec:3_science_analysis}
Spectroscopic redshifts determined for the galaxies in our $5\sigma$
line list, as well as photometry in the CDFS field from the FIREWORKS
catalogue (\citealt{Wuyts2008}; $U_{38}$, $B_{435}$, $B$, $V$,
$V_{606}$, $R$, $i_{775}$, $I$, $z_{850}$, $J$, $H$,
$K_{s}$, [3.6], [4.5], [8.0]), and in the FIRES field (\citealt{Forster2006}; $U$, $B$, $V$,
$V_{606}$, $I_{814}$, ${J}_{s}$, $H$, and
${K}_{s}$), served as inputs to the stellar
PEGASE.2 \citep{fioc1997} population models, as described in
\citet{Glazebrook2004}. The models were fit to aperture magnitudes, and
the final stellar masses\footnote{PEGASE.2 only accounts for
  luminous stars in its determination of stellar mass. Thus stellar
  remnants such as white dwarfs, neutron stars, and black holes were
  not included.} were subsequently scaled according to the total $K_s$ magnitude.

Emission line luminosities were first converted to ``fiducial'' star formation
rates ($SFR_f$), starting from the \citet{Kennicutt1998} relation
\begin{equation}
	SFR_f (M_{\sun}/yr) = \left[\frac{10^{0.4 A_{H_{\alpha}}}}{0.5}\right]\left[\frac{7.9\times10^{-42}}{1.82}\right]\left[ L_{[OII]}(ergs/s) \right],
	\label{eq:SFR_LOII_conversion}
\end{equation}
where the factor of 1.82 accounted for the conversion from a Salpeter
IMF \citep{Salpeter1955} to the BG03 IMF \citep{Baldry2003}, and we
assume $A_{H\alpha}=1$.  In practice, $A_{H\alpha}$ and the other
coefficients in Equation~\ref{eq:SFR_LOII_conversion} are likely to
depend strongly on the galaxy stellar mass.  We therefore use the
empirical correction advocated by \citet[Eq.8]{Gilbank2010}, based on
analysis of the SDSS: 
\begin{equation}
	SFR (M_{\sun} yr^{-1}) = \frac{SFR_f}{\left\lbrace(a)\cdot \tanh{\left[(X+b)/c\right]}+d \right\rbrace},
	\label{eq:SFR_emp_corr}
\end{equation}
where $X=\log{\left(M_{*}/M_{\sun}\right)}$, $a=-1.424$, $b=-9.827$, $c=0.572$, and
$d=1.700$.  
We apply this correction to all SFR reported in this paper.

This SFR estimate assumes that the \oii\ emission arises from gas ionized by massive stars; any contribution from an active galactic nucleus (AGN) would reduce the SFR.  We would not expect AGN to contribute significantly in such low-mass galaxies; in \citet{Gilbank2010_ROLESII} we confirmed from analysis of mid-infrared colours and {\it Chandra} point sources that AGN were entirely negligible in that $z=1$ sample.  We will therefore also neglect AGN contamination, except when we include higher mass galaxies from other surveys (\S~\ref{sec:fors2_extension}).
\subsection{Extension of the survey to higher masses}
\label{sec:fors2_extension}
\begin{figure}
	\includegraphics[width=1.0\linewidth]{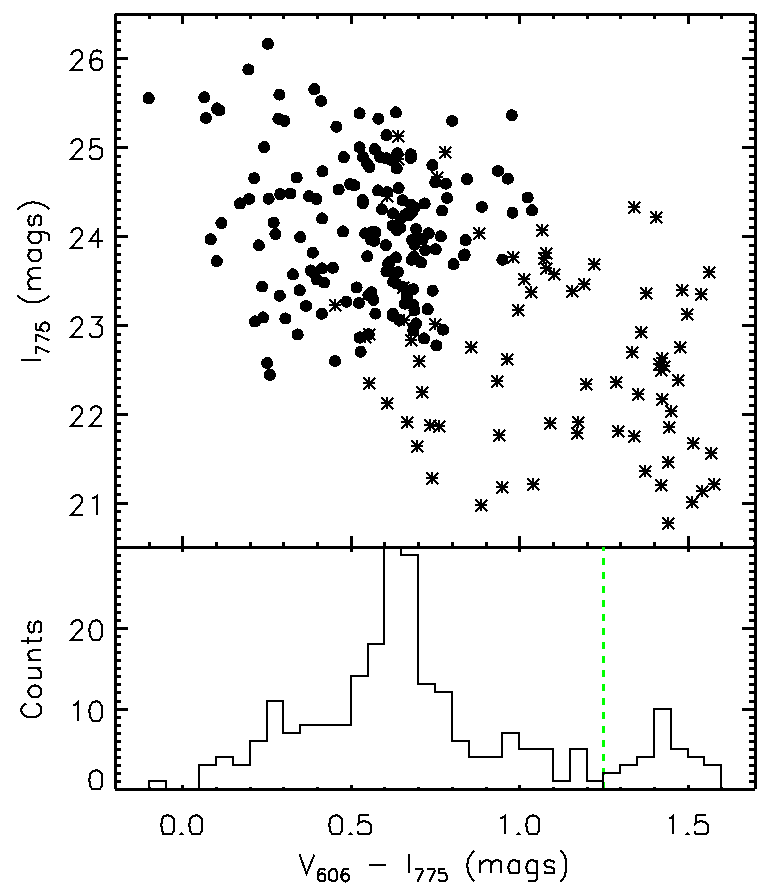}
	\caption{The colour-magnitude diagram of the CDFS and FORS2
          targets is shown. {\it Upper
          Panel: }  Our targets in the CDFS are shown as solid black circles while
          the public FORS2 data are represented by black
          asterisks. {\it Bottom Panel: }
          The overall distribution of $V_{606}-I_{775}$ colour is shown.  The red sequence
          cut is shown as the vertical green dashed line; we exclude
          galaxies redder than this limit from the remainder of our analysis.} 
	\label{fig:red_seq_cmd}
\end{figure}
We supplement our survey with brighter (more massive) galaxies,
primarily from publicly available VLT/FORS2
spectroscopy overlapping our CDFS sample area, from \citep{Vanzella2008}.  Their sample was a colour and photometric redshift selected
catalogue with targets found between the redshift ranges of $0.5 \lesssim
z \lesssim 2$ and $3.5 \lesssim z \lesssim 6.3$. We selected only those
targets which were found within the LDSS-3 field-of-view centered on
our CDFS field pointings, and which fell within our redshift range of
$0.62 < z < 0.885$. Their observation masks used 1\arcsec\ slits (compared to
0.8'' for ROLES) and exposure times for each mask were typically $\ge
4$ hours. Hereafter this higher mass sample is referred to as
\textit{FORS2}. 

For these massive galaxies, there is more concern that the emission
lines could arise from AGN or LINER emission, rather than star
formation.  As in ROLES, we therefore exclude red-sequence galaxies
from the sample, using a colour-magnitude diagram (CMD)
consisting of bands which bracket the $4000\mbox{\AA}$ break at
$z\sim0.75$. Figure \ref{fig:red_seq_cmd} shows the characteristic
bimodal colour distribution of galaxies in our CDFS and FORS2 samples; while all our low-mass targets
are in the blue cloud, a subset of the FORS2 galaxies
are on the red sequence, with $V_{606}-I_{775}~>~1.25~\mbox{mag}$.

For each FORS2 emission line
we calculated $V_{max}$ as before (with a typical value of
$4.7~\times~10^4~\mbox{ Mpc}^3$). 
The $K$-magnitude binned FORS2 completeness was
determined in the same way as for our data, and the results are shown
in Figure \ref{fig:ROLES_cdfs_fors2_completeness}. We extracted spectroscopic redshifts, line identifications and quality
flags, and 1D spectra for these galaxies.  [OII] emission line fluxes
were measured  
from the 1D spectra in the same way as for our own data. A constant $4\sigma$
noise flux limit of $6~\times~10^{-18} \mbox{erg s}^{-1} \mbox{cm}^{-2}$ was adopted,
approximately the same as the average noise flux limit for the rest of
our sample (see
Figure \ref{fig:noise_flux_culled_dets}). 

\subsection{Completeness}
\label{sec:completeness}
\begin{figure}
	\includegraphics[width=1.0\linewidth]{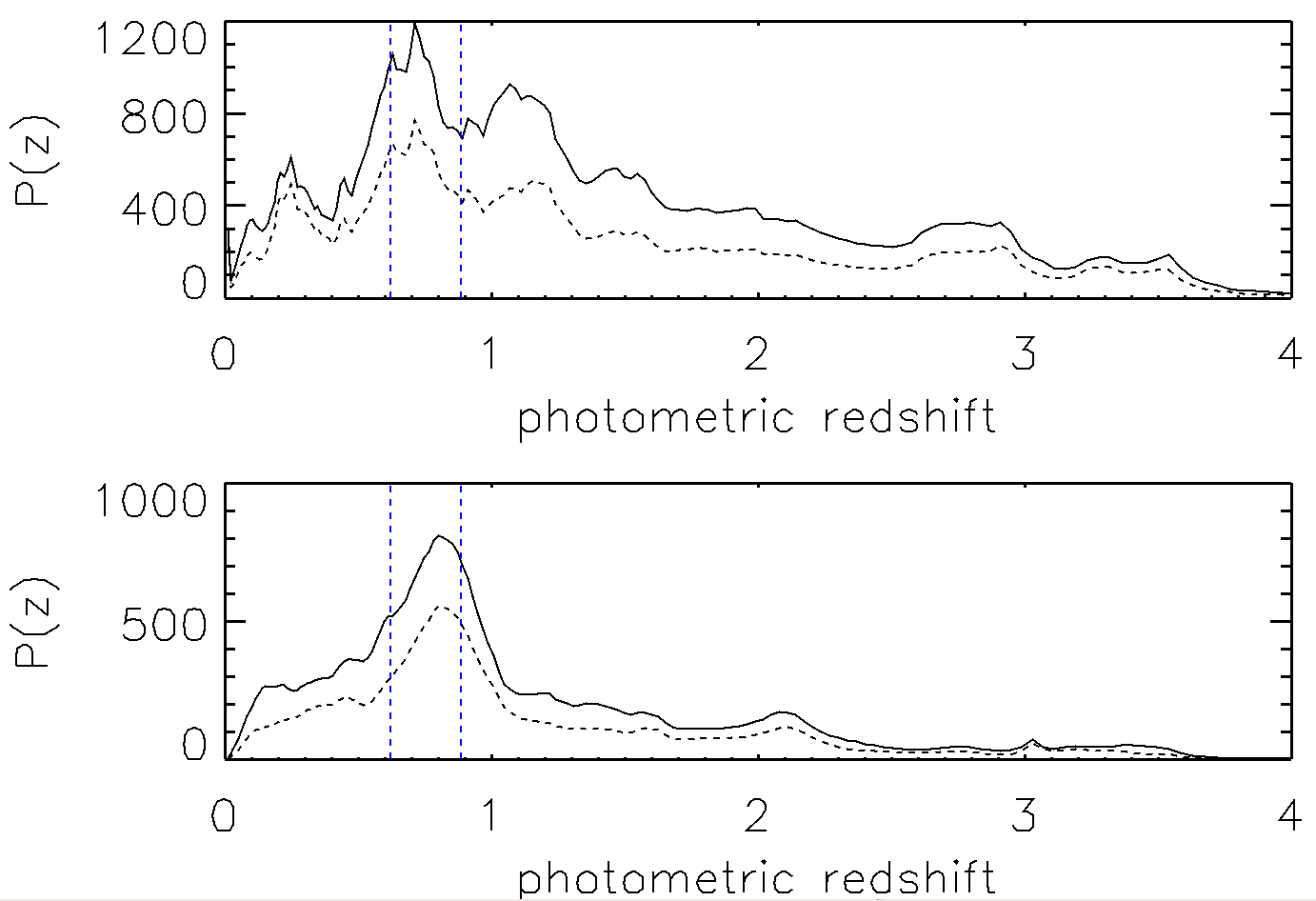}
	\caption{{\it Upper Panel: } Summation of all photometric
          redshift PDFs within the LDSS-3 FOV and with $22.5 < K < 24$
          (black curve) compared with the summation of those
          photometric redshift PDFs corresponding to galaxies targeted
          in this survey (lower green curve), for the CDFS field. {\it
            Lower Panel: }Same as the upper panel, corresponding to the FIRES field.} 
	\label{fig:ROLES_K_mag_completeness_redshift}
\end{figure}

As for ROLES, we characterize our spectroscopic completeness as
follows.
For each field, all photometric redshift probability
distribution functions corresponding to galaxies within our target
fields were first binned by $K$-magnitude.  Within each bin the PDFs were
summed, giving the total redshift distribution for all
galaxies in each bin, $P_K(z)$. The summed redshift distribution in
each bin was then integrated over the redshift of interest here, $0.62 < z < 0.885$. 
The process was then repeated for only those galaxies that were
successfully targeted (i.e., the slit was successfully extracted), and
the ratio of this to the total distribution yields the targeting
completeness.  The redshift PDFs of targeted and candidate galaxies are
shown in Figure~\ref{fig:ROLES_K_mag_completeness_redshift}.

\begin{figure}
	\includegraphics[width=1.0\linewidth]{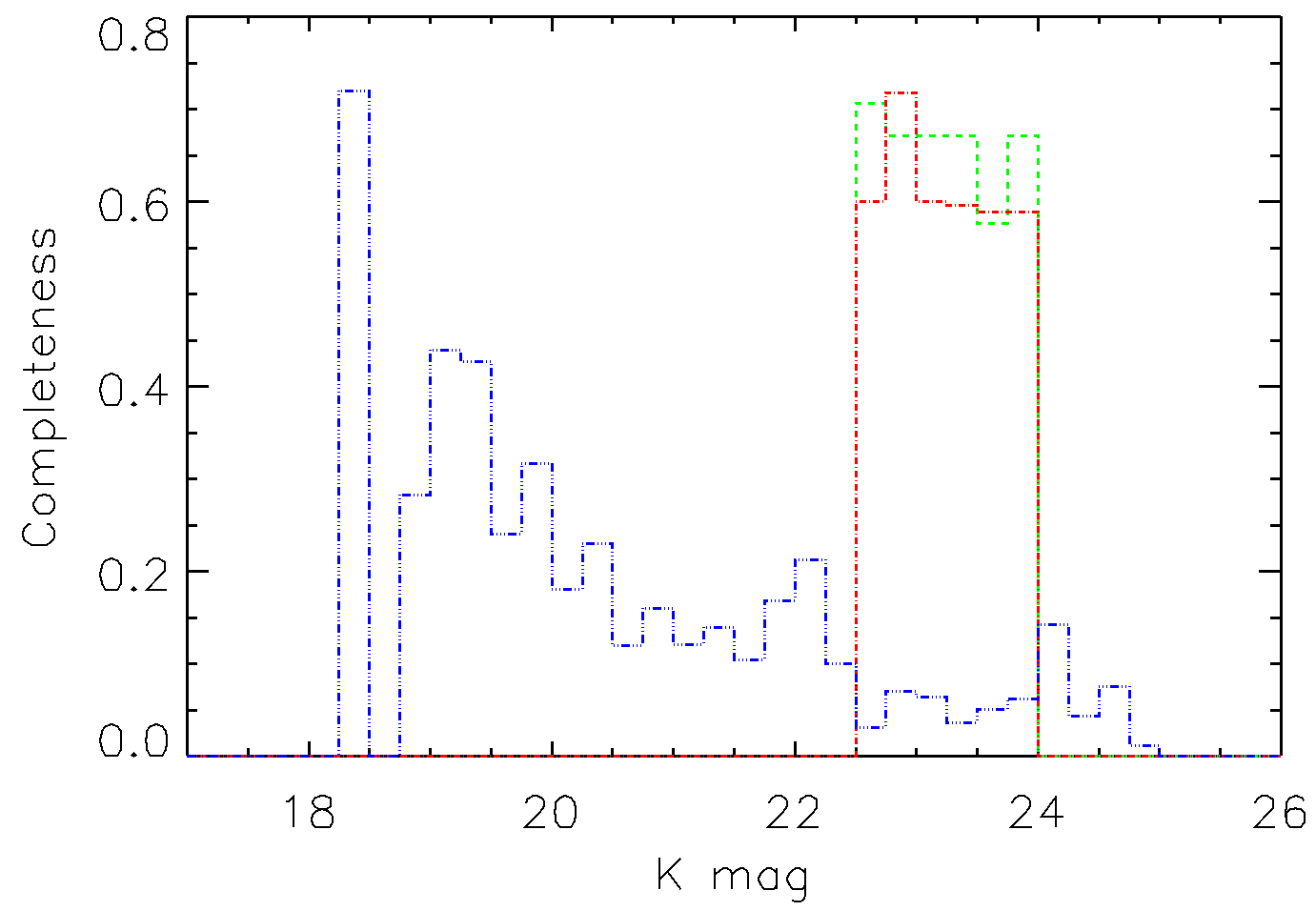}
	\caption{Completeness histograms for our surveys in the CDFS
          and FIRES are shown as the red and green lines.  We also
          include existing FORS2 data for brighter galaxies, shown as
          the blue histogram.  }  
	\label{fig:ROLES_cdfs_fors2_completeness}
\end{figure}

The resulting completeness is high, $\sim 70$ per cent, and independent
of $K$ magnitude.  This is shown in Figure \ref{fig:ROLES_cdfs_fors2_completeness} with the CDFS field
represented by the red dashed line and the FIRES field denoted by the
green dashed line.  The figure also includes spectra for brighter
galaxies from the public domain, discussed in \S~\ref{sec:fors2_extension}.

\subsection{Redshift distribution and definition of dense environments}\label{sec-struct}

\begin{figure}
	\includegraphics[width=1.0\linewidth]{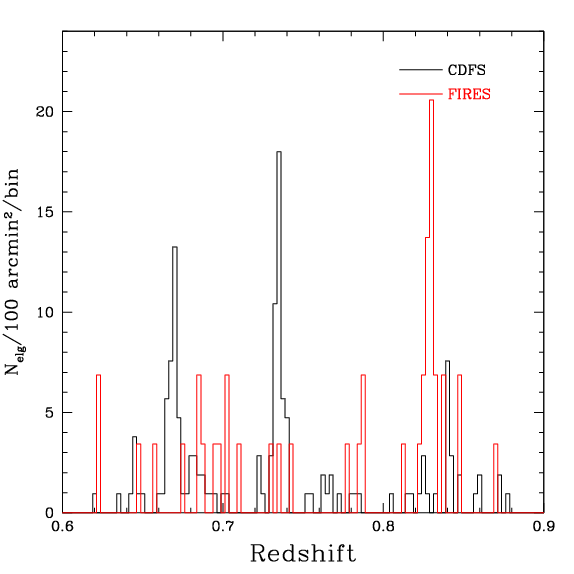}
	\caption{The redshift distribution of emission line galaxies in our
          sample.  The distributions are normalized to an area of 100 arcmin$^2$, close to the area of the CDFS and $\sim 3.6$ times larger than the FIRES field.
Large-scale structure, due to the known walls in the
        CDFS and the cluster MS1054-03 in the FIRES field, is identifiable as three narrow redshift peaks.}
	\label{fig:structure_redshift_distrib}
\end{figure}

The redshift distribution of our emission line galaxies is shown 
in Figure \ref{fig:structure_redshift_distrib}.   Two prominent peaks in this distribution
at $z\sim 0.68$ and $z\sim 0.73$ correspond to the well-known ``wall-like'' structures  in
the CDFS \citep{Gilli2003,LeFevre2004,Vanzella2005,Ravikumar2006}.  We
associate all CDFS galaxies with $|z-0.668|<0.016$ and
$|z-0.735|<0.009$ with these structures.  As traced by our low-mass galaxy sample, this structure is spread fairly uniformly over the LDSS3 field of view, without an apparent density gradient or central concentration.  The rest-frame velocity dispersion of emission line galaxies in each of these structures is $970$km/s and $430$km/s, respectively.
The
other important structure, at $z\sim 0.83$, is the MS1054-03 cluster,
in the FIRES field \citep{Forster2006}; all galaxies in the field and within $\Delta
z=0.02$ of this redshift are associated with the cluster.  The rest-frame velocity dispersion of these emission line galaxies is $1300$km/s, in good agreement with \citet{vD+00}.
Together,
these three subsets of galaxies are referred to as ``dense
environments'' for the subsequent analysis.  Combined, the subsample comprises
 112 galaxies, 23 of which are associated with the MS1054-03 cluster.  The remaining galaxies are referred to as the ``field''; it now represents an underdense sample relative to the average of our full sample.  We will show in \S~\ref{sec:mass_function} that both the CDFS ``walls'' and the MS1054 cluster are comparably overdense, by a factor of at least $7$ relative to the field, and probably more like a factor $\gtrsim 45$.

\subsection{[OII] Luminosity, Stellar mass and SFR Limits}
\label{sec:luminosity_mass_SFR_limits}
\begin{figure}
	\includegraphics[width=1.0\linewidth]{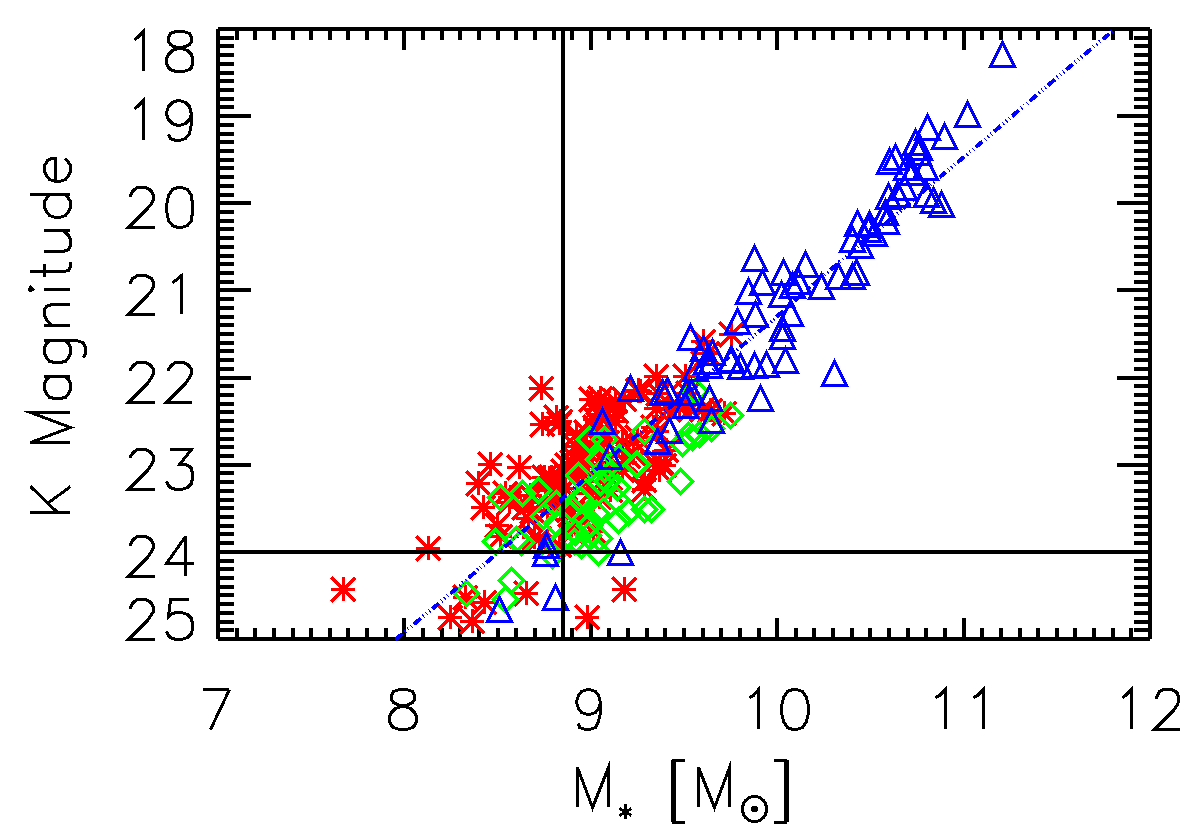}
	\caption{$K$-magnitude as a function of stellar mass for emission
          line galaxies in our two survey fields (red for CDFS, and
          green for FIRES), supplemented with the more massive
          FORS2 galaxies (blue triangles).  
          The solid lines show the $K$-magnitude and
          adopted stellar mass limits, while the blue 
          line is the best fit to the three populations
          combined.} 
	\label{fig:k_mag_stellar_mass} 
\end{figure}

The stellar mass limit of our sample is determined from the scatter in the
correlation between $K$-magnitude
and stellar mass shown in Figure \ref{fig:k_mag_stellar_mass}. The
horizontal line shows our limiting selection magnitude of $K =
24$. Based on the scatter in this relation, the sample is nearly ($>90$ per cent)
complete in stellar mass for $\log{(M_\ast/\Msun)}\gtrsim 8.85$. We take this to be our $2\sigma$ mass completeness limit;
the sample extends to lower masses, but is
systematically missing galaxies with high $M/L_K$ ratios.

\begin{figure}
	\includegraphics[width=1.0\linewidth]{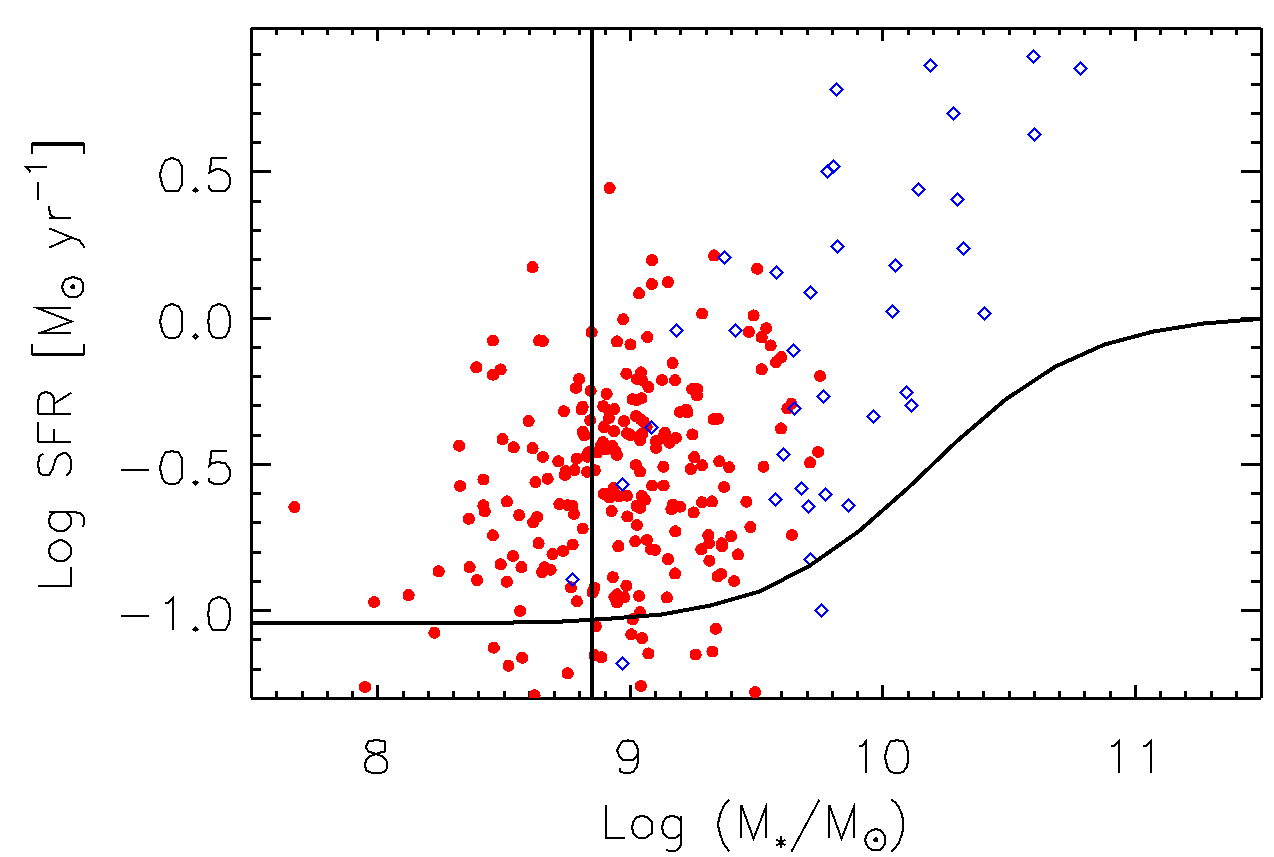}
	\caption{The SFR of all emission-line galaxies in our sample is shown as a function of their
          stellar mass. Our LDSS3 data are shown as the red points;
          blue points are public data from FORS2 spectroscopy.  The $2\sigma$ stellar mass and $5\sigma$ SFR limits are shown as
          solid, black lines. }
	\label{fig:sfr_stellar_mass}
\end{figure}

The average $5\sigma$ [OII] flux limit as a function of wavelength was shown in Figure
\ref{fig:noise_flux_culled_dets}.  The sample is statistically complete
for fluxes as low as $\sim5\times10^{-18} \mbox{ergs~s}^{-1}\mbox{ cm}^{-2}$.
Considering our low
redshift bound of $z\sim0.62$, the corresponding [OII] luminosity limit
at which the sample should be statistically complete is
determined to be $\log{L_{[OII]}}\sim39.9$.  However, most of the volume
is limited to higher luminosities, and $\log{L_{[OII]}}\sim40.1$ is a
more representative limit for most of the data.
The limiting SFR (including the mass-dependent empirical correction) can be determined from the 
[OII] luminosity limit, 
using Equations \ref{eq:SFR_LOII_conversion}
and \ref{eq:SFR_emp_corr}. Figure \ref{fig:sfr_stellar_mass} is a plot
of the empirically corrected SFRs versus stellar
mass. The $2\sigma$ mass and $5\sigma$ SFR limits are indicated with solid lines.
We reach SFR$~\gtrsim 0.1\Msun$ at the low stellar masses of interest
here, corresponding to a mass doubling time (assuming a recycling
factor $R=0.5$) of $t_d\sim1.3\times10^{10}$ years.  This is almost twice the
Hubble time at $z=0.7$, and we expect this depth is sufficient to
capture most of the star formation at these masses \citep[e.g.][]{Noeske2008,Gilbank2011}.  

\subsection{Survey Volume and density estimates}
\label{sec:survey_volume}
Our survey volume is defined by the survey area, the limiting magnitudes
($22.5 < K < 24$), and a flux limit on the emission line detection.  
As in ROLES, we determine the redshift limit for which the galaxy would
fall outside our $K$ magnitude limits, including the $k$-correction term
\begin{equation}
	k_{corr}(z) = \frac{-2.58z + 6.67z^2 - 5.73z^3 - 0.42z^4}{1 - 2.36z + 3.82z^2 - 3.53z^3 + 3.35z^4}.
	\label{eq:K_corr}
\end{equation}

The wavelength-dependent flux limit shown in
Figure~\ref{fig:noise_flux_culled_dets} was similarly used to determine
the redshift limits over which each detected emission line would be
observable.
The volume, $V_{max}$, from which a galaxy with a detected emission
line could have been found was then calculated from the survey area (105.62 and 29.15 square
arcminutes for our CDFS and FIRES pointings, respectively) and the redshift space bounded
by the $K$-magnitude and noise flux limits, and the wavelength limits of
our spectra. 
The volume, $V_{max}$, was determined for each galaxy by integrating the 
differential co-moving volume (see \citealt{Hogg1999}) between the
appropriate redshift limits.
The maximum ($0.62 < z < 0.885$) $V_{max}$ is $4.7~\times~10^4~\mbox{ Mpc}^3$
in the CDFS fields, and $1.3~\times~10^4~\mbox{ Mpc}^3$ in FIRES.

The three structures are defined by the redshift limits given in \S~\ref{sec-struct}.  Interpreting the redshift limits as cosmological, removing these regions reduces the volume of our field sample by $\sim 20$ per cent.  For the MS1054 cluster and CDFS walls themselves, which are decoupled from the Hubble flow, most of the galaxies are likely located within a volume that is much smaller than this cosmological volume.  We will assume their line-of-sight extent is 10 Mpc (i.e. assuming a $\sim 5$ Mpc virial radius, which is still probably conservatively large).  This corresponds to a volume of $\sim 200$Mpc$^{-3}$ for the cluster, and $\sim 520$Mpc$^{-3}$ and $\sim 610$Mpc$^{-3}$ for the two CDFS walls.

\subsection{ROLES and SDSS Stripe 82 data}\label{sec-comp}
We will compare our results with similarly-selected data, from ROLES
\citep{Gilbank2010_ROLESII} at $z\sim1$, and the local
Universe from our Stripe82 analysis \citep{Gilbank2011}.  Both of
these samples are consistent with our present analysis in the choice of IMF, the empirical
calibration of [OII] to SFR, and the removal of red, massive galaxies\footnote{In fact, the Stripe 82 analysis is based on
H$\alpha$ measurements of the SFR which, by construction, are consistent with 
the mass-dependent \oii-SFR conversion.  This avoids the incompleteness in \oii~ noted by \citet{Gilbank2011} for high mass galaxies in Stripe 82.}. 

The ROLES SFR are also limited by \oii\ flux, and thus the mass-dependent SFR limit has the same form as shown in Figure~\ref{fig:sfr_stellar_mass}.  However, the greater luminosity distance and brighter sky at the wavelength of redshifted \oii\ at $z=1$ means, despite the longer exposure times, that the limiting SFR at $z=1$ is about a factor $\sim 3$ greater than in the present study.  In \citet{Gilbank2011}, we demonstrated that, locally, the SFRD has converged for $\mbox{SFR}>0.1 M_\odot$/yr, for galaxies with $\log{\left(M/M_\odot\right)}\sim 9$, and $\mbox{SFR}>1 M_\odot$/yr for galaxies with $\log{\left(M\right)}\sim ~10$.   Globally, the average sSFR is known to evolve approximately as $(1+z)^{2.5}$ \citep[e.g.][]{PBJ}.   Assuming the shape of the SFRD does not evolve strongly, we would expect ROLES ($\log{\left(M/M_\odot\right)}\gtrsim 9$ at $z=1$) to be complete at $\sim 0.5M_\odot$/yr, and the present study ($\log{\left(M/M_\odot\right)}\gtrsim 9$ at $z=0.75$) to be complete at $\mbox{SFR}\sim 0.4 \mbox{M}_\odot/\mbox{yr}$.  Thus, we expect our samples are deep enough to have recovered most of the star formation in the Universe, and to be fairly insensitive to the precise choice of limiting SFR.

\section{Results}
\label{sec:4_results}

\subsection{Star formation rate density}
\label{sec:sfrd}

\begin{figure}
        \includegraphics[width=1.0\linewidth]{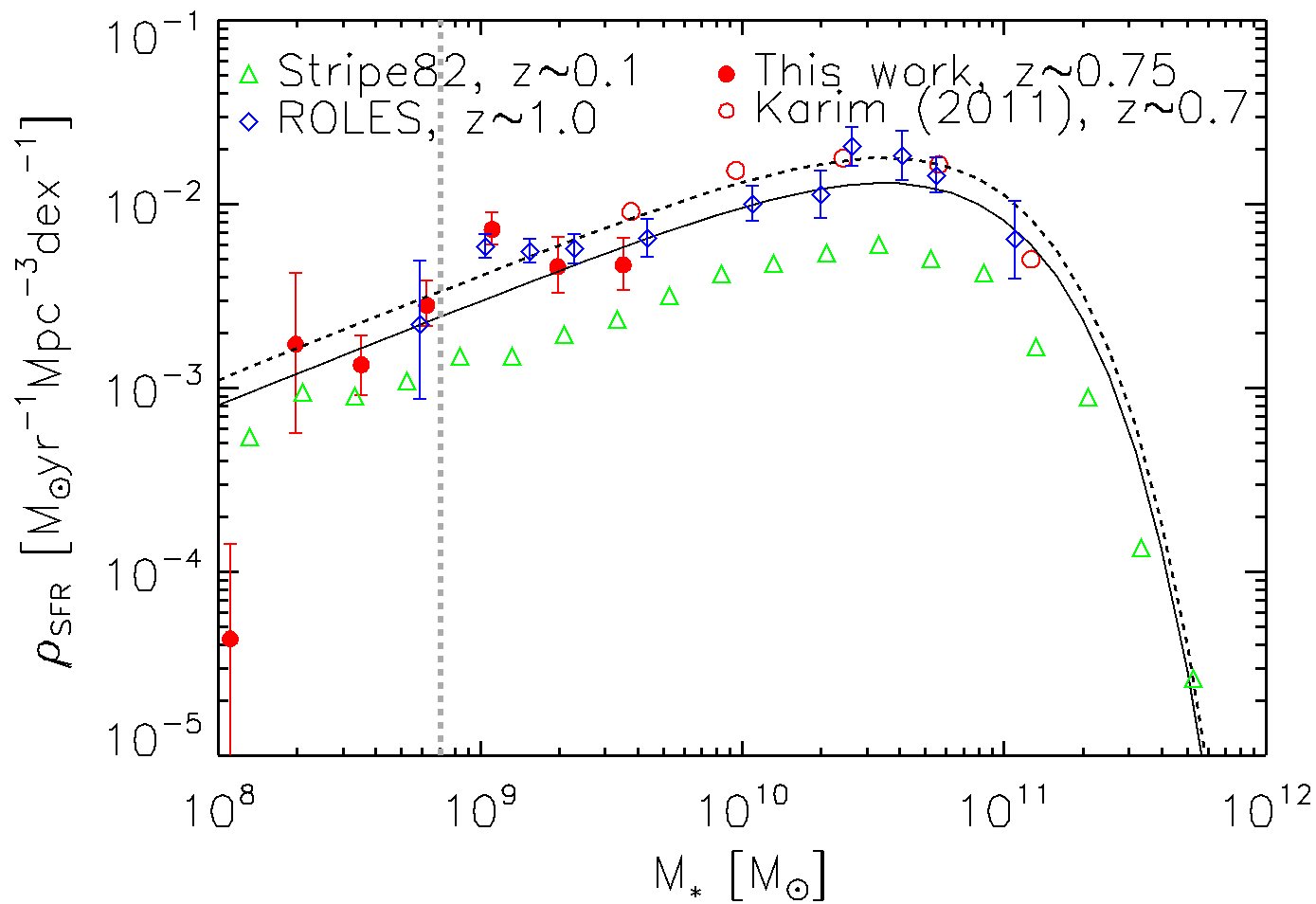}
        \caption{The SFRD of our sample, at $z~\sim 0.75$, is shown as
          the red points with error bars at $M<3\times 10^9 M_\odot$.    The grey vertical dashed line highlights the
          mass limit of the present survey.  The
          high-mass data are taken from the stacked radio
          analysis of galaxies at $0.6<z<0.8$ from \citet{Karim}, shown
          as the open circles.  Note that
          \citet{Karim} find a factor $\sim 2$ difference in
          normalization between their radio-based SFRD and the [OII]-based analysis
          of ROLES at $z=1$.  If this is treated as a systematic effect, the solid
          circles of \citet{Karim} on this Figure should be brought down by a factor
          two.  This SFRD is compared 
          with ROLES at $z=1$ (blue diamonds), and the SDSS from our analysis of
          Stripe82 data (green triangles).  The solid and dashed line represent 
          Schechter functions fit to the Stripe82 data, scaled by a factor $(1+z)^2$ to $z=0.75$ and $z=1$.  }
        \label{fig:sfrd}
\end{figure}

The SFRD was computed as follows:
\begin{equation}
        \rho_{SFR}(M_\ast) = \displaystyle \sum\limits_{i=0}^n \frac{P_{OII,i} \cdot SFR_{i}}{w_i \cdot V_{max,i}},
        \label{eq:SFRD}
\end{equation}
where the SFR was calculated according to Equations
\ref{eq:SFR_LOII_conversion} and \ref{eq:SFR_emp_corr}, and the sum is
over all galaxies in a given bin of stellar mass.  $P_{OII,i}$ is the
probability that the line is [OII], relative to the probability of it
being any other plausible emission line; $w_i$ is a magnitude-dependent
weight to account for incompleteness.  
Figure \ref{fig:sfrd} shows the SFRD for our LDSS3 sample.
Recall from
Figure~\ref{fig:ROLES_cdfs_fors2_completeness} that our completeness drops significantly for bright ($K<22.5$) galaxies, corresponding to $M>3\times
10^9M_\odot$.  
Moreover, this high mass end has no contribution from the FIRES field, and is therefore especially
sensitive to cosmic variance (not included in the error bars) given the
large scale structure in the CDFS field.  Finally, no aperture corrections
have been applied, which means SFRs are likely
underestimated for the most massive
galaxies. Thus we limit our data to $M<3\times 10^9 M_\odot$, and replace the higher mass measurements with the $0.6<z<0.8$ data of \citet{Karim}, who used stacked radio
observations in the COSMOS field to measure SFR over a wider field than
we have here.  They note that there is a factor $\sim 2$ difference in
normalization between [OII]--based SFR and their analysis.  To eliminate this likely systematic offset,  the \citet{Karim} SFRD should be reduced (or our 
[OII] data should be increased) by this
factor.  However, we show the data as measured, with no rescaling applied.

We compare this with results from ROLES
\citep{Gilbank2010_ROLESII} at $z\sim1$, and the local
Universe from our Stripe82 analysis \citep{Gilbank2011}. 
Interestingly, the ROLES SFRD is consistent with our new data at $z\sim 0.75$,
over most of the mass range, with some evidence for higher SFR at
$M>10^{10}M_\odot$ if the SFRD from \citet{Karim} are reduced by a factor
$\sim 2$ as described above.  Compared with the
$z\sim 0.1$ data from Stripe 82, the SFRD at both $z=0.75$ and $z=1$ is higher by a
factor $\sim 3$ at all masses. Given the statistical and systematic
uncertainties, though, we cannot rule out a 
continuous evolution of the form $SFRD\propto(1+z)^{2.0}$ \citep[e.g.][]{Gilbank2010_ROLESII}, represented
on Figure~\ref{fig:sfrd} as the solid and dashed lines at $z=0.75$ and $z=1$, respectively.

Our findings are in good agreement with \citet{Karim} and
\citet{Gilbank2010_ROLESII}, that there is no evidence here
for strong evolution in the shape of the SFRD.  We note that both ROLES
and the present work hint at an increased contribution from the
lowest-mass galaxies for which we are complete ($\sim 10^9M_\odot$),
relative to the local Universe.  However, any mass-dependent effect
is fairly subtle, and it is unlikely that systematic uncertainties are
understood to this level across the entire mass and redshift range.
The data are not inconsistent with a simple, mass-independent evolution
of $SFRD\propto(1+z)^{2.0}$ over the redshift range $0<z<1$, as
suggested by \citet{Gilbank2010_ROLESII}.

\subsection{Stellar mass function of star-forming galaxies}
\label{sec:mass_function} 
\begin{figure*}
       { \includegraphics[width=0.48\linewidth]{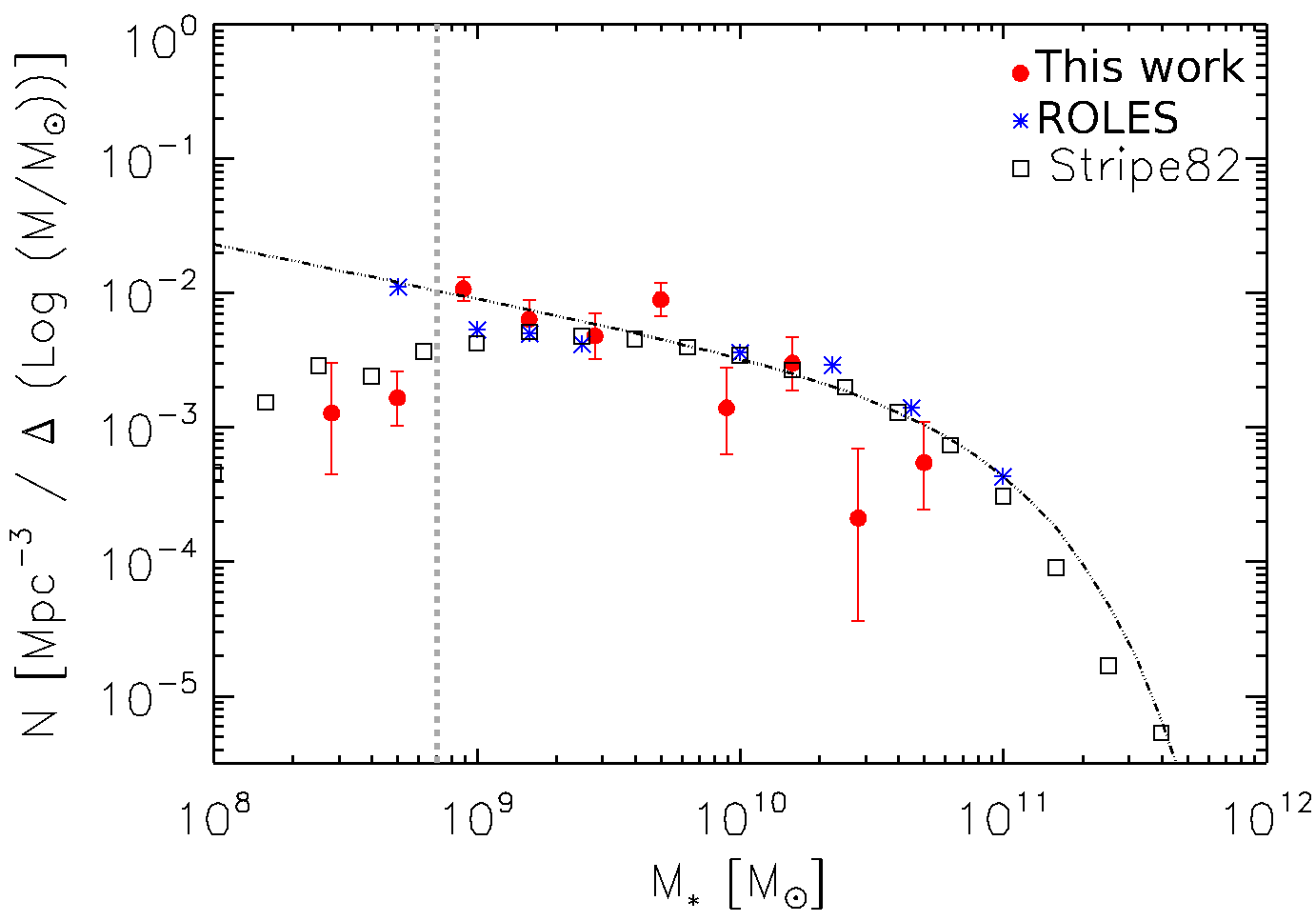} 
       \includegraphics[width=0.48\linewidth]{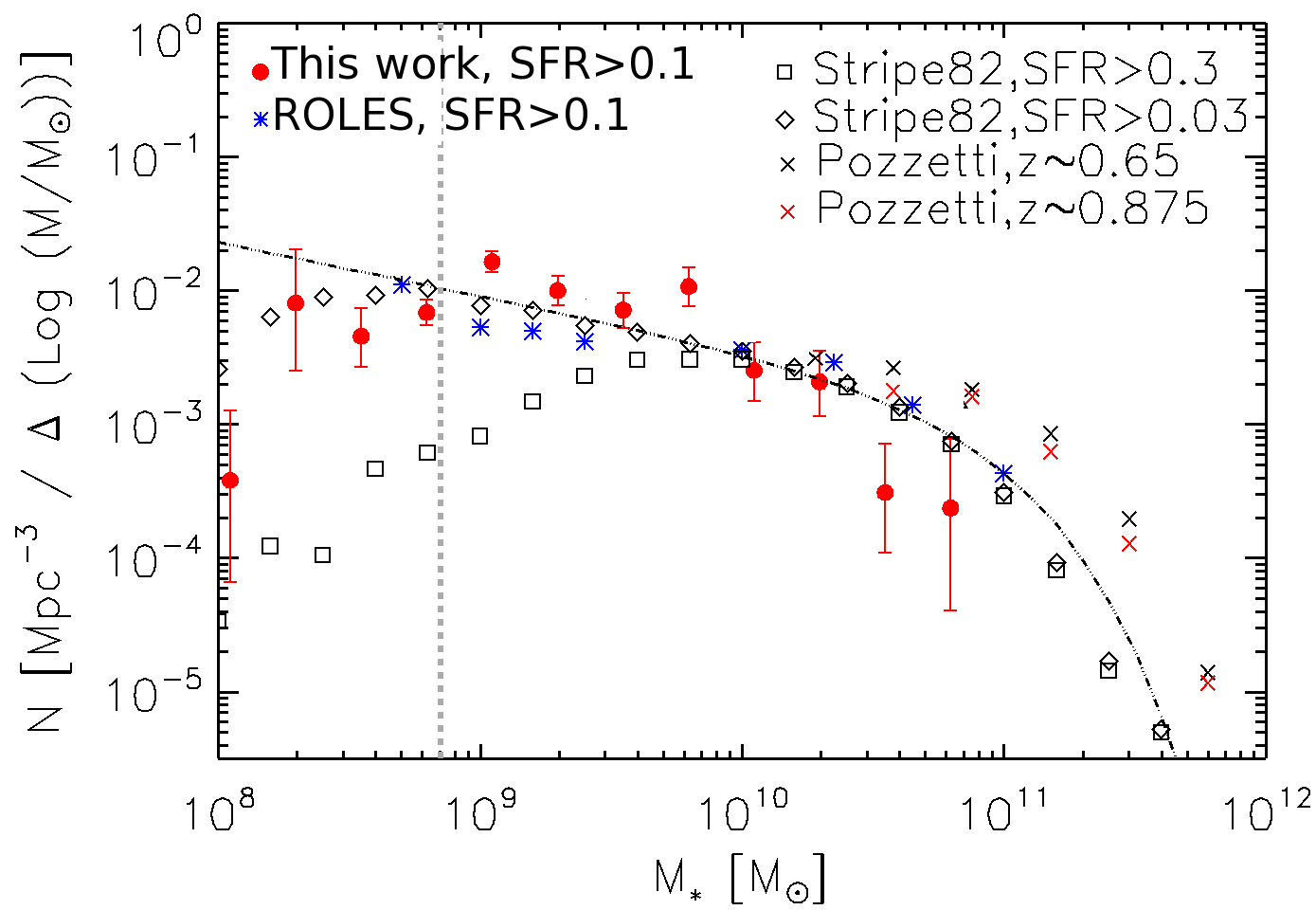}}
\caption{ {\it Left panel: }The stellar mass function of 
          all star--forming galaxies with SFR$\ge$0.33\msunyr\ in our sample is shown as the
          points with error bars.  This is compared with ROLES data at
          $z\sim 1$ (blue asterisks), to the same SFR limit, and with
          the SDSS Stripe 82 data to a limit of a factor $\sim 3$ lower to
          account for evolution \citep[see][]{Gilbank2010_ROLESII}.  The dashed line represents the Schechter function fit to the local, SDSS data by \citet{Peng2010}.  
          {\it Right panel:}  Our data are now shown to a deeper SFR
          limit of  SFR$\ge$0.1\msunyr; this is compared to the
          Stripe82 data to two different depths, as described in the
          legend.  We also compare with data at $0.55<z<0.75$ and
          $0.75<z<1.0$ from \citet{Pozzetti2009}, shown as black and red crosses
          respectively.  
The
          thick grey vertical dashed line in both panels highlights the mass limit of
          the present survey.}
        \label{fig:mass_function_1}
\end{figure*}

The galaxy stellar mass function for our sample of star-forming galaxies
(SF-GSMF) is shown in Figure \ref{fig:mass_function_1}. 
In order to study its evolution, we compare our
z$\sim$0.75 data with the Stripe 82
SDSS data as detailed in \S~\ref{sec-comp}.   The left
panel only considers the present z$\sim$0.75 data above a conservative SFR limit
of 0.3 \msunyr~so as to compare directly with the depth of the slightly
shallower z$\sim$1 ROLES data. To compare fairly with the
local data, the Stripe 82 sample is limited to $SFR\geq 0.1$
\msunyr.  Since the SFRD falls by a factor of $\approx$3 over this
redshift interval (as discussed in  \S~\ref{sec:sfrd}), this cut corresponds to a
similarly evolving limit.  For comparison, we also show the single Schechter function fit to the SDSS star-forming population at $z=0.1$ from \citet{Peng2010}.  Transforming to the \citet{Baldry2003} IMF, the parameters of this fit are $\log {M_\ast}=10.92$ and 
  $\Phi^*=2.612\times10^{-3} \mbox{dex}^{-1} \mbox{Mpc}^{-3}$.  
The high-mass end of the SF-GSMF function ($M\gtrsim3\times10^{9}M_\odot$) is
not very well-determined in our present data, due to the limited survey
area and spectroscopic sampling.  At lower masses,
where we expect our sample to be highly complete, the data are
consistent with no evolution of the
SF-GSMF from $0<z<1$, in agreement with the conclusions of \citet{Gilbank2010_ROLESII} and 
\citet{Peng2010}.  

In the right panel, the full depth of the present
z$\sim$0.75 dataset is considered, down to a SFR limit of 0.1\msunyr.  The
Stripe 82 data from the left panel is reproduced here, but we also show
these data extended to a lower limiting SFR$\ge$0.03\msunyr, allowing
for a factor 
$\approx$3 in SFR evolution.
We also compare with data from 
 \citet{Pozzetti2009}, shown without error bars, for clarity.
This sample defined star--forming
galaxies as those with $\log{sSFR/\mbox{Gyr}}>-1$, approximately consistent
with the limits we apply to our data, here.  
Again we convert the result to correspond to a BG03
\citep{Baldry2003} IMF.  The
  results are consistent with little or no evolution in the SF-GSMF
  over this redshift range.  However, it also demonstrates the sensitivity
of any measured evolution to the limiting SFR of the samples.  If we
account for the global evolution of SFR we find the SF-GSMF remains
constant down to the lowest masses for which we have statistically
complete samples.  Choosing a fixed, non-evolving limit would result in
a large decrease with increasing cosmic time of the SF-GSMF at the low
mass end.  The difference between the analysis of \citet{Peng2010},
where SF galaxies are identified strictly by colour, and our Stripe82
analysis, shows that remaining systematic uncertainties of this type
are at least as important as any physical evolution.

\begin{figure}
        \includegraphics[width=1.0\linewidth]{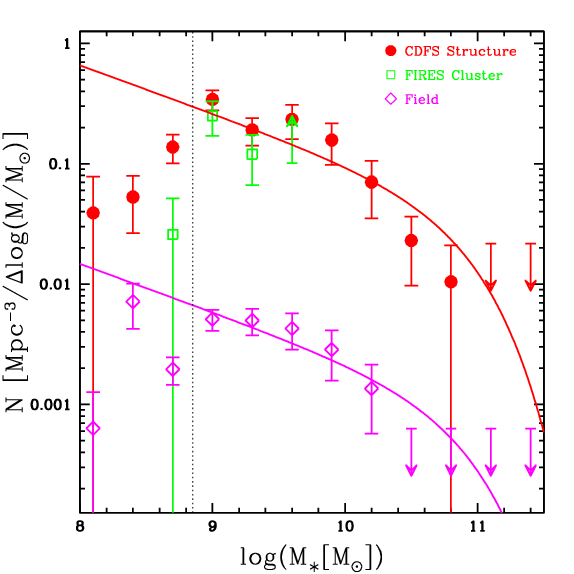}
        \caption{The stellar mass function for star--forming galaxies
          in our survey is shown,  divided by environment as shown in the legend. Here we assume that the line-of-sight extent of both the cluster and the CDFS structures is 10 Mpc (comoving).  The local Schechter function for star-forming galaxies from 
\citet{Peng2010} has been renormalized to fit the low-density and structured environments.  We see no dependence of the shape on environment.  Both the cluster and the CDFS ``walls'' are comparable in overdensity, a factor $\sim 45$ times denser than the field.  Error bars and upper limits are $1\sigma$.  The lower limit on the most massive FIRES point is due to the lack of redshifts for $K<22$.
The
          vertical dotted line highlights the mass limit of
          the present survey.
          } 
        \label{fig:mass_function_2}
\end{figure}

In Figure~\ref{fig:mass_function_2} we divide our sample into different
environments: the MS1054 cluster in FIRES, the large-scale structure in CDFS,
and the remaining population which we call the ``field''.  The cluster in FIRES has no public spectroscopy for $K<22$, which means that field is incomplete for $M>10^{9.5} M_\odot$.  Recall from \S~\ref{sec:survey_volume} that the normalizations of the FIRES cluster and CDFS structures are calculated assuming a comoving line-of-sight extent of $10$ Mpc.   We also show the local mass function of star-forming galaxies, from \citet{Peng2010}, renormalized to minimize the $\chi^2$ value of the dense and underdense samples over the range for which the data are complete.  This shows that both high density environments are a factor $\sim 45$ times denser than the field sample.  This depends on our assumption about the line-of-sight extent.  A firm lower limit to the overdensity is a factor $7$, obtained by assuming the cosmological volume between the redshift limits used to define each sample. We note that the shape of the single Schechter function, with $M_\ast$ and $\alpha$ fixed to their local values, is a good fit to both the field and overdense samples.  There is no evidence for it to vary with environment, and the reduced $\chi^2$ is near unity for both samples.

\subsection{Specific star formation rate}
\label{sec:ssfr}
\begin{figure}
        \includegraphics[width=1.0\linewidth]{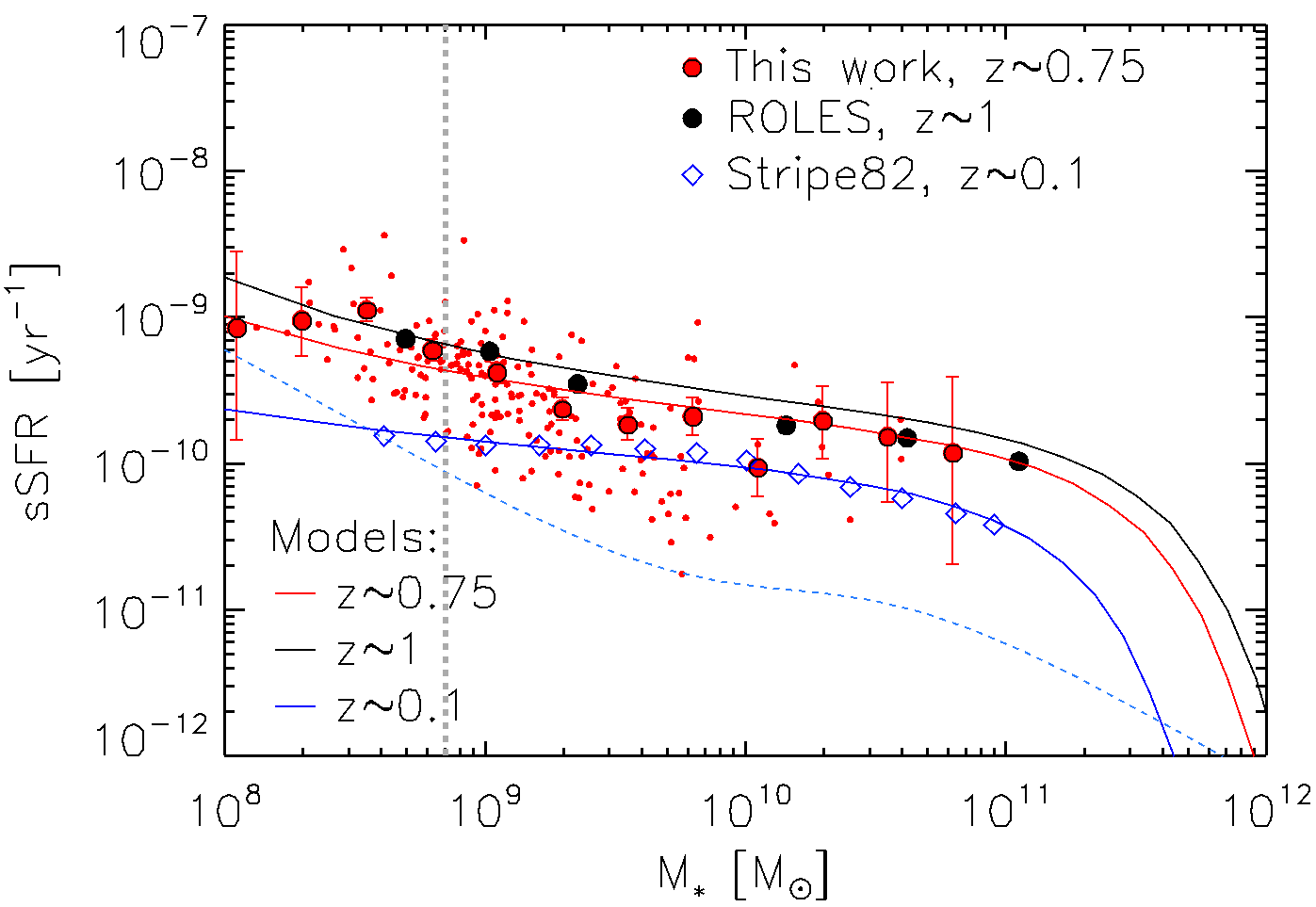}
        \caption{The sSFR of the star-forming galaxies in our sample
          are shown as a function of stellar mass.  Small points
          represent individual galaxies, while the filled circles with
          error bars are the mean values, in equal mass bins.
          Comparable mean values are shown at $z=1$ from ROLES (solid black circles),
          and at $0<z<0.1$ from Stripe82 \citep{Gilbank2011}, 
 without error bars for clarity.  We overplot the \citet{Gilbank2011} models (see text for description) at three redshifts, as indicated in the legend.  Also shown is the mass limit of this survey (thick grey vertical dashed line), and the SFR limit (dashed, light blue line).} 
        \label{fig:ssfr_1}
\end{figure}

The sSFR of galaxies in our sample is shown as
a function of their stellar mass
in Figure \ref{fig:ssfr_1}.  The small filled
circles represent the individual emission line galaxies. The large symbols represent the \textit{binned mean}
sSFR of the present, combined sample (solid red circles), compared with
ROLES at $z=1$ (solid black circles), and local star-forming SDSS
(blue diamonds) datasets.  
\begin{figure}
        \includegraphics[width=1.0\linewidth]{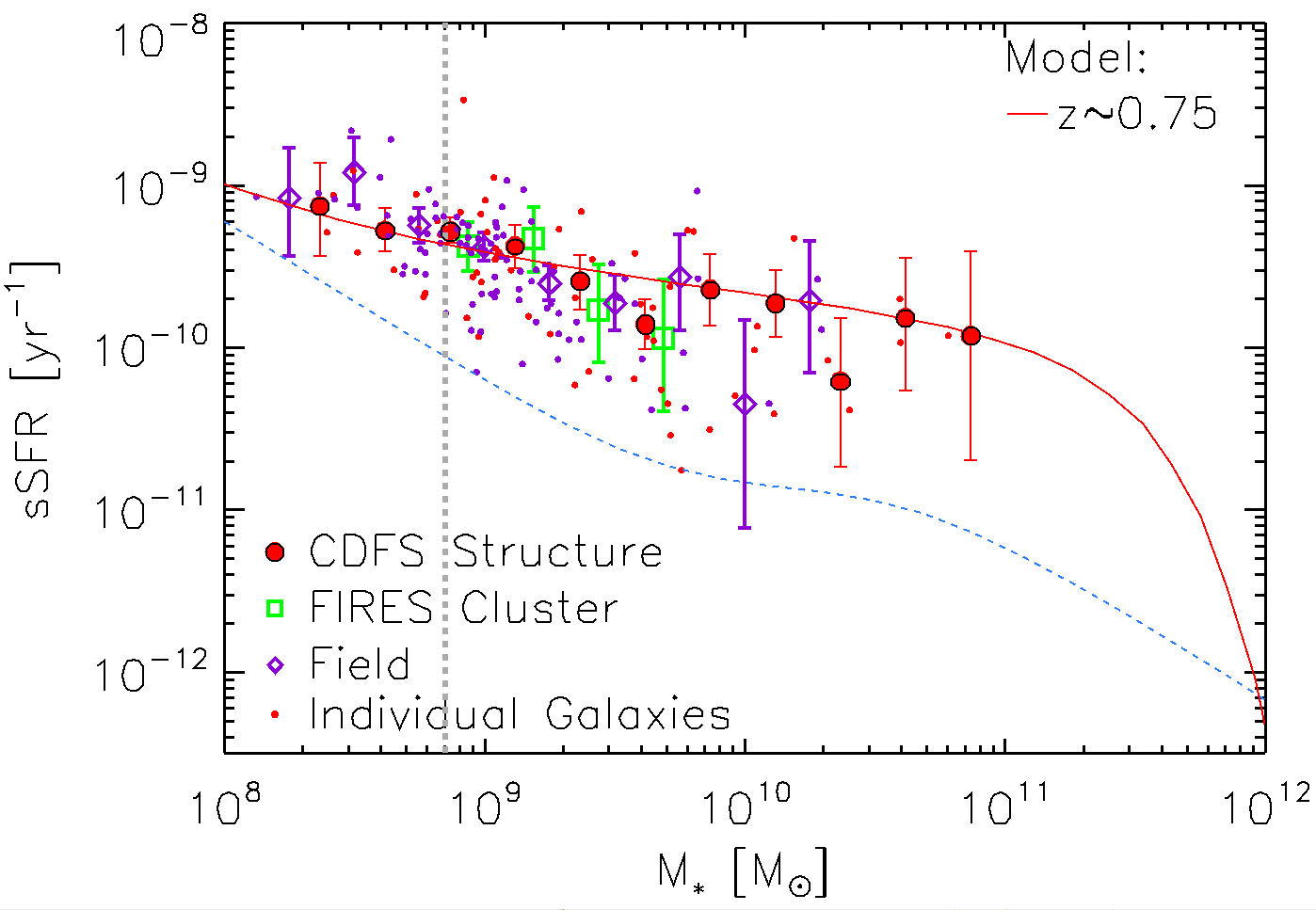}
        \caption{As Figure~\ref{fig:ssfr_1}, but where our data are
          now divided into high density structures, the MS1054 cluster
          (open green squares), and field environments. We show only the \citet{Gilbank2011} model corresponding to $z=0.75$, as the smooth, red curve.}
        \label{fig:ssfr_2}
\end{figure}

We find a distinct anti-correlation between $sSFR$ and $M_\ast$, with a power-law slope of $\beta \sim -0.2$ over the mass range $10^9<M_\ast/M_\odot<10^{10}$.  This is similar to our finding at $z=1$ with ROLES, and steeper than the anti-correlation at $z=0$ which already poses a challenge to models \citep{BBC,Weinmann12}.   
At all stellar masses, the sSFR at $z=0.75$ and $z=1$ is significantly
larger than locally.  Moreover, the relation significantly departs from a simple power-law, with an 
upturn observed at low stellar mass \citep[see also ][]{Brinchmann2004,Elbaz2007,Popesso2011}.  
To interpret this, we turn to the models used by \citet{Gilbank2011}, based on the staged galaxy formation models of \citet{Noeske2008}.  These are shown as the smooth curves on Figure~\ref{fig:ssfr_1}.  In this model, galaxies are parametrised by an exponentially declining SFR, with formation redshift and SFR timescales both a function of stellar mass.  This simple description provides a reasonable match to the observations at all three epochs shown, including the increase in sSFR observed at the lowest stellar masses in both the present data and ROLES.

Finally, in Figure \ref{fig:ssfr_2} we show the $\mbox{sSFR-}M_\ast$ relation in different environments.
Both high and typical density populations show a
decreasing sSFR with increasing stellar
mass. The shape and normalization of the relation in all environments
are consistent with one another, and with the model of \citet{Gilbank2011}, over the entire stellar mass range.

\section{Discussion}
\label{sec:5_discussion}

It has consistently been shown that the main influence of environment
is on the fraction of star-forming galaxies of a given stellar mass \citep[e.g.][]{Baldry2006,2012arXiv1205.3368W}.
Amongst the star-forming population itself, any residual environment
dependence is weak at best \citep[e.g.][]{Peng2010}.  However, a
transformation from active to passive cannot be instantaneous and must
therefore be reflected in an environmental dependence of the sSFR
distribution.  Results here are more controversial and depend on how
the analysis was done.  At low redshifts, $z\lesssim 0.5$, dense
environments consistently show either no change in the sSFR relation
\citep[e.g.][]{Peng2010} or a mildly reduced sSFR \citep[e.g.][]{V+10}.  At $z\sim 1$ and above, there is
some evidence that sSFR is {\it higher} amongst star forming galaxies
in some dense environments, but not all
\citep{Sobral2011}.  While we will not resolve this issue here, our
data extends the discussion to low-mass galaxies at an intermediate
redshift of $z\sim 0.75$.  

\begin{figure}
	\includegraphics[width=1.0\linewidth]{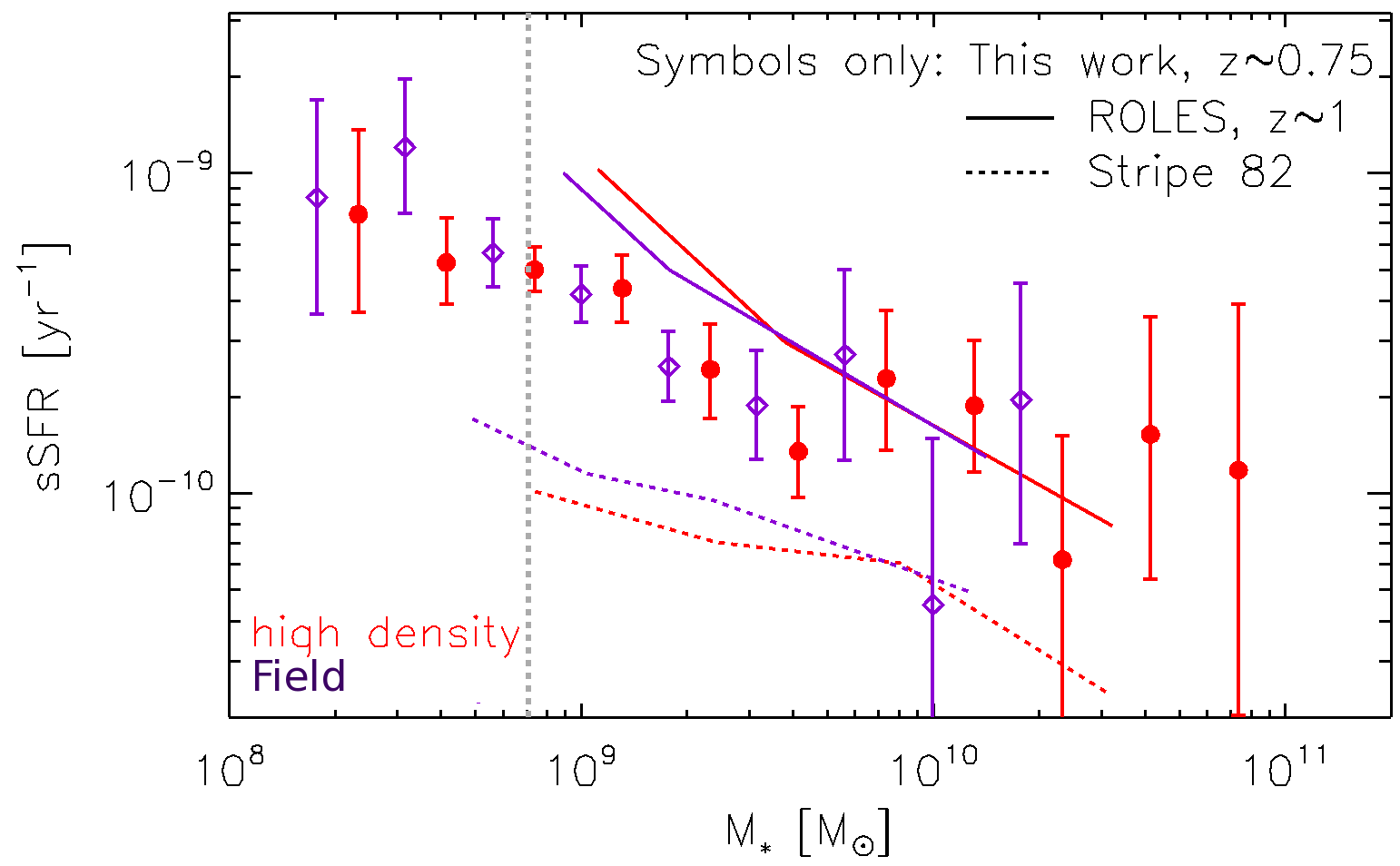}
	\caption{The mean correlation between sSFR and stellar mass is shown for our sample, as the points with error bars.  The purple diamonds represent the result in the ``field'', while the red circles represent the high-density environments (MS1054 cluster and CDFS structures), that are a factor $\sim 45$ overdense.  This is compared with similar relations at
          $z=1$ (ROLES) and $z=0$ (SDSS), taken from the analysis by
          \citet{Li2011}. In this case, the red line only represents an overdensity of $\sim 6$ relative to the average environment, represented by the purple line. The thick grey vertical dashed line 
          highlights the mass limit of the survey.} 
	\label{fig:ssfr_tornado_density}
\end{figure}

In Figure \ref{fig:ssfr_tornado_density} we re-create the density
segregated, $sSFR-M_\ast$ plot from \citet[][their Figure 6]{Li2011}. In
this plot we show the density-dependent $sSFR-M_\ast$ relation for each of
the $z=0.1$ (SDSS), $z=0.75$ (present study) and $z=1$ (ROLES) epochs. 
The SDSS and ROLES
samples were segregated according to a local density parameter, $\rho_5$,
which is defined in detail\footnote{Briefly, $\rho_5$
indicates the redshift-completeness weighted number density of
star-forming galaxies limited to $M_{K_{AB}} \le -21.0$, found within a
`nearest-neighbour' volume defined by the five closest galaxies to the
current galaxy being evaluated. The volume is defined by the maximum
projected radius of the set of five nearest-neighbour galaxies, and the
difference in co-moving distances set by the \textit{closest} and
\textit{farthest} nearest-neighbour galaxies in
redshift space. } in \citet{Li2011}. In the Figure, we show only the relations for the average environment, and the most overdense subsample, for best comparison with the present data.
For the ROLES sample at $z=1$, \citet{Li2011} find the local density ranges from 0.04 Mpc$^{3}$ to 0.6 Mpc$^{-3}$, with an average of $\sim 0.1$  Mpc$^{3}$.  Thus their overdense regions are a factor $\sim 6$ denser than the average.   We estimated the volume density of our subsamples in \S~\ref{sec-struct}; assuming a $10$ Mpc line-of-sight extent for our high density structures this is $\sim 0.9$ Mpc$^{-3}$, compared with the field value of $\sim 0.02$ Mpc$^{-3}$.  While these numbers cannot be compared directly with $\rho_5$, they are not too dissimilar in practice.  The small fields mean that the nearest-neighbour approach taken by \citet{Li2011} is almost exclusively identifying structure in redshift space, as we do here.  The main difference is that Li et al. use massive galaxies as tracers, while we use low-mass emission line galaxies.  While the physical scales over which the density is estimated will not be identical, we may expect that the relative density between structures and the field can be fairly compared between the two analyses, to within a factor of a few.  Thus, the high-density environments of the present study, overdense by a factor $\sim 45$, are likely to be significantly denser than those of \citet{Li2011}.  In fact, $45$ is probably an underestimate, as at least the MS1054 virial diameter is likely much less than $10$ Mpc.  

Considering first the average, ``field'' environments, Figure~\ref{fig:ssfr_tornado_density} shows smooth evolution of the $sSFR-M_\ast$ relation from $z=0$ to $z=1$, with little or no significant change in slope, but a normalization that increases approximately as $(1+z)^{2.5}$.  In contrast, the low-mass end slope of the relation in high-density environments shows mild evolution.  It is flatter than the average relation locally, and steeper than the average relation at $z=1$.  Interestingly, our new data at an intermediate redshift $z\sim 0.7$ shows no difference at all between the two environments.  Since our survey includes environments with much higher densities, the actual evolution between $z\sim 0.7$ and $z\sim 1.0$ of comparably dense environments could be considerably stronger than shown here.  Note that the analysis of \citet{Li2011} includes an {\it underdense} environment, for which the contrast with their densest environments is considerably larger.  

The implication is that the sensitivity of low-mass, star forming
galaxies to their environment has evolved significantly from $z=1$ to
the present day. Today, the average sSFR of such galaxies is slightly
lower in high-density environments, while at $z=1$ the average is
slightly {\it higher}.  This ``reversal'' of the SFR-environment
relation has been noted by others \citep[e.g.][]{Elbaz2007,Li2011}, and
our new data at $z=0.7$ appear to correspond to the ``transition''
epoch where the sSFR--mass relation shows no environmental dependence. 

\section{Conclusions}
\label{sec:6_conclusions}
For the first time, the faint [OII]~emission from low stellar mass galaxies
($8.5~<~\log{(M_\ast/\Msun)~<~9.5)}$ has been spectroscopically measured for
galaxies in the redshift range $0.62~<~z~<~1.15$.   By targetting fields (CDFS and FIRES) with known overdensities, including the massive MS1054-03 cluster, we explore how star formation in these low-mass galaxies are affected by their environment.
Our main conclusions are as follows:
\begin{itemize}
\item There is little, if any evolution in the galaxy stellar mass
function of [OII] luminous galaxies between $z=0$ and $z\sim 1$.  
\item The trend of a decreasing specific star formation rate with
increasing stellar mass has been confirmed down to unprecedented
stellar masses. The normalization at $z\sim 0.75$ is similar to our
earlier results at $z=1$, and significantly higher than at $z=0$.
\item The average power-lase of the sSFR$-M_\ast$ relation is $\beta\sim -0.2$, with indication of a steeper relation at low masses.  This is consistent with what we found at $z=1$ with ROLES.
\item The star formation rate density shows little evolution between $z=0.7$ and
  $z=1$, but is consistent with the $(1+z)^{2.0}$ evolution expected from
  comparison with the SDSS.  The SFRD evolution is consistent with a
  mass-independent evolution in normalization; that is, the
  characteristic mass of star--forming galaxies is independent of redshift.  However we caution that
  systematic and statistical uncertainties preclude us from
  establishing the SFRD to better than a factor of $\sim 2$ at any mass;
  thus there is room for relatively small, mass-dependent evolution.
\item Environment is found not to influence the $sSFR-M_\ast$ relationship
at any stellar mass at the epoch studied here,
$z\sim0.75$. This suggests that the SFR of star-forming
galaxies is not enhanced or diminished by local density, and that the
apparent reversal in correlation between environment and star formation
rate occurs at higher redshift.
\end{itemize}

Our results on the constancy of the SF-GSMF are consistent with many
previous works \citep[e.g.][]{Pozzetti2009,Gilbank2011}, extending the results to lower masses at this
intermediate redshift.  In \citet{Li2011} we
showed the emergence of a weak environmental dependence of the
sSFR--mass relation on environment, such that the lowest mass galaxies
in the densest environments show a significant excess sSFR compared to
their lower density counterparts. In this work, we show the absence of
this trend $\approx$1 Gyr later.  By the present day, another $5$ Gyr later, the relation has reversed, and low-mass galaxies in dense environments have lower sSFR than the average.  This smooth transition is consistent with e.g. \citet{Quadri2012}, who pointed out that it would be strange to see a sharp
transition given factors such as the apparently smooth growth of
passive galaxies over cosmic time.  We note, however, that this environmental-dependence of the star--forming population is very mild, and much weaker than the aforementioned evolution in the fraction of galaxies with no star formation whatsoever.

\section*{Acknowledgments}
\label{sec:acknowledgements}
We thank an anonymous referee for very helpful comments that improved the paper.
The ROLES collaboration would like to thank S. Crawford, T. Dahlen,
H. Dominguez, M. Franx, S. Juneau, C. Maier, B. Mobasher, L. Pozzetti,
E. Vanzella, and S. Wuyts for providing data and useful correspondence
throughout the project.  MLB acknowledges support from an NSERC
Discovery Grant, and an Early Researcher Award from the Province of Ontario.
\label{lastpage}
\bibliographystyle{apj} 
\bibliography{ROLES-650_resubmit}
\end{document}